\documentclass[twocolumn,aps,prl,superscriptaddress,floatfix]{revtex4}
\usepackage[T1]{fontenc}
\usepackage[utf8]{inputenc}
\usepackage[english]{babel}
\usepackage{color}
\usepackage{float}
\usepackage{braket}
\usepackage{amsmath}
\usepackage{amstext}
\usepackage{amssymb}
\usepackage{graphicx}
\usepackage{bbold}
\usepackage[unicode=true,pdfusetitle,
bookmarks=true,bookmarksnumbered=false,bookmarksopen=false,
breaklinks=false,pdfborder={0 0 1},backref=false,colorlinks=false]
{hyperref}
\hypersetup{
	colorlinks,linkcolor=blue,citecolor=blue,urlcolor=blue}
\usepackage{soul}
\makeatletter
\date{\today}

\makeatother
\setul{0.5ex}{0.3ex}
\definecolor{Blue}{rgb}{0,0.0,1}
\setulcolor{Blue}
\usepackage{babel}

\begin{document}

\author{Luis M. Canonico}
\affiliation{Instituto de F\'\i sica, Universidade Federal Fluminense, 24210-346 Niter\'oi RJ, Brazil}
\author{Tarik P. Cysne}
\affiliation{Instituto de F\'\i sica, Universidade Federal do Rio de Janeiro, Caixa
	Postal 68528, 21941-972 Rio de Janeiro RJ, Brazil}
\affiliation{Departamento de F\'\i sica, Universidade Federal de S\~ao Carlos, Rod. Washington Lu\'\i s, km 235 - SP-310, 13565-905 S\~ao Carlos, SP, Brazil }
\author{Tatiana G. Rappoport}
\affiliation{Instituto de F\'\i sica, Universidade Federal do Rio de Janeiro, Caixa
	Postal 68528, 21941-972 Rio de Janeiro RJ, Brazil}
\affiliation{Department of Physics and Center of Physics, University of Minho, 4710-057, Braga, Portugal}	

\author{R. B. Muniz}
\affiliation{Instituto de F\'\i sica, Universidade Federal Fluminense, 24210-346 Niter\'oi RJ, Brazil}

\title{Two-dimensional orbital Hall insulators}

\begin{abstract}
	
Detailed analyses of the spin and orbital conductivities are performed for different topological phases of certain classes of two-dimensional (2D) multiorbital materials. Our calculations show the existence of orbital-Hall effect (OHE) in topological insulators, with values that exceed those obtained for the spin-Hall effect (SHE). Notably, we have found non-topological insulating phases that exhibit OHE in the absence of SHE. We demonstrate that the OHE in these systems is deeply linked to exotic momentum-space orbital textures that are triggered by an intrinsic Dresselhaus-type of interaction that arises from a combination of orbital attributes and lattice symmetry. Our results strongly indicates that other classes of systems with non-trivial orbital textures and/or orbital magnetism may also exhibit large OHE even in their normal insulating phases.
		
	\end{abstract}
\maketitle

	The OHE, similarly to the SHE, refers to the creation of a transverse flow of orbital angular momentum that is induced by a longitudinally applied electric field~\cite{Orbitronics-Bernevig}. It has been explored mostly in three dimensional metallic systems, where it can be quite strong~\cite{Inoue-OHEdTransitionMetal,Orbitals-Frustration-Inoue,OrbitalGiantSHE-Inoue,Go-OHETexture}.  For systems in which the spin-orbit coupling (SOC) is sizeable, the orbital and spin angular momentum degrees of freedom are coupled, establishing an interrelationship between charge, spin, and orbital angular momentum excitations. However, the OHE does not necessarily require SOC, it can be associated to the presence of orbital textures~\cite{Go-OHETexture} and be especially significant in various materials.

Chiral orbital textures in the reciprocal space have been discussed in connection with orbital magnetism at the surface of $sp$ metals~\cite{Go2017}, photonic graphene~\cite{Nalitov2015} and also in topological insulators with strong SOC.  More recently they were observed in chiral borophene~\cite{Miwa2019},  single-layer transition metal dichalcogenides ~\cite{Chen2019} and tin telluride monolayers for photocurrent generation~\cite{Kim2019}.  Orbital magnetism is enhanced in surfaces~\cite{surface}, indicating that orbital effects can be crucial in 2D materials, which can also be evidenced by the observation of orbital textures in van der Waals materials. Still, OHE remains mostly unexplored in 2D materials \cite{GraphaneOHE, Phong-Mele-2019}. 
	
Here, we investigate the role of orbital textures for the OHE displayed by multi-orbital 2D materials. We predict the appearance of rather large OHE in these systems both in their metallic and insulating phases. The orbital Hall currents can be considerably larger than the spin Hall ones, and be present even in the absence of SHE. Their use as information carriers widens the development possibilities of novel spin-orbitronic devices.

In our analyses, we consider a minimal tight-binding (TB) model Hamiltonian that involves only two orbitals ($p_x$ and $p_y$) per atom in a honeycomb lattice~\cite{ReisBismutheneExperimental,antimonene}:

\begin{equation}
{\cal H}=\sum_{\langle i j\rangle} \sum_{\mu \nu s} t_{i j}^{\mu \nu}{p^\dagger_{i \mu s}}p_{j \nu s}+ \sum_{i \mu s}\left(\epsilon_{i} + \lambda_I \ell^z_{\mu \mu}\sigma^z_{s s}\right) p^\dagger_{i \mu s} p_{i \mu s},
\label{eqn:HNN}
\end{equation}
\noindent  where $i$ and $j$ denote the honeycomb lattice sites positioned at $\vec{R}_i$ and $\vec{R}_j$, respectively. The symbol $\langle i j \rangle$ indicates that the sum is restricted to the nearest neighbour (n.n) sites only. The operator $p^{\dagger}_{i \mu s}$ creates an electron of spin $s$ in the atomic orbitals $p_\mu=p_{\pm}=\frac{1}{\sqrt{2}}(p_x\pm ip_y)$ centred at $\vec{R}_i$. Here, $s=\,\uparrow,\downarrow$ labels the two electronic spin states,  and $\epsilon_i$ is the atomic energy at site $i$, which may symbolise a staggered on-site potential that takes values $\epsilon_i = \pm V_{AB}$, when site i belongs to the A and B sub-lattices of the honeycomb arrangement, respectively. The transfer integrals $t_{i j}^{\mu \nu}$ between the $p_{\mu}$ orbitals centred on n.n atoms are parametrised according to the standard Slater-Koster TB formalism~\cite{Slater-Koster}. They depend on the direction cosines of the n.n. interatomic directions, and may be approximately expressed as linear combinations of two other integrals ($V_{pp\sigma}$ and  $V_{pp\pi}$) involving the $p_{\sigma}$ and $p_{\pi}$ orbitals, where $\sigma$ and $\pi$ refer to the usual components of the angular momentum around these axes. 

Since our model does not include the orbital $p_z$, it is restricted to a sector of the $\ell=1$ angular momentum vector space spanned only by the eigenstates  of $\ell^z$ $\big|p_\pm\big>$ associated with $m_\ell=\pm1$. Within this sector it is useful to introduce a pseudo angular momentum $SU(2)$-algebra where the Pauli matrices act on $\big|p_\pm\big>$. In this case, there is a one-to-one correspondence between the representations of the Cartesian components of the orbital angular momentum operators in this basis and the usual Pauli matrices, and $\ell^z$ is not conserved (details are given Sec. I of the supplementary material - SM). The last term in Eq.\ref{eqn:HNN} describes the intrinsic atomic SOC. 

\begin{figure}[h]
	\centering\includegraphics[width=1.\linewidth,clip]{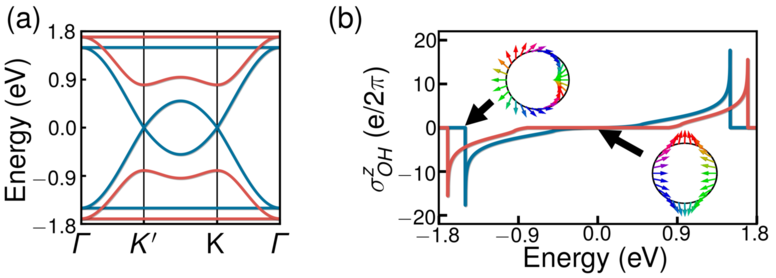}
	
	\caption{(a) Band structure calculations along some symmetry lines in the 2D BZ for $V_{pp\pi}=0$, $V_{pp\sigma}=$1 eV, and $\lambda_I = 0$. The blue line represents the results for $V_{AB}=0.0$, and the red line for $V_{AB}=0.8$. 
	(b) Orbital Hall conductivities calculated for the same sets of parameters. The insets show the in-plane contribution to the orbital angular momentum textures calculated in the neighbourhoods the $\Gamma$ (left inset) and $K$ (right inset) symmetry points of the 2D Brillouin zone, for $V_{AB}=0.0$. The left and right inset textures are associated with the lower flat and dispersive bands, respectively.
	}
	\label{fig:textures} 
\end{figure}

This simple model describes relatively well the low-energy electronic properties of novel group V based 2D materials~\cite{li2018new,zhou2018giant,antimonene}. Its topological characteristics were previously investigated in the context of optical lattices, and it has been verified that it exhibits a rich topological phase diagram, which includes quantum spin-Hall insulator (QSHI) phases~\cite{flat-bands-Wu,px-py-WU,MultiorbitalHoneycomb-Wu,li2018new,PRLPxPy-Nosotros}.

Following Ref. \onlinecite{MultiorbitalHoneycomb-Wu} we shall assume, for simplicity, that $V_{pp\pi}=0$ and $V_{pp\sigma}=1$ eV.  Our focus is on three distinct phases that manifest themselves depending on the parameters specified in Eq. (\ref{eqn:HNN}). In the absence of SOC and sub-lattice resolved potentials, the electronic band structure consists of four gapless bulk energy bands, two of which form Dirac cones at the $K$ and $K'$ symmetry points of the 2D first Brillouin zone (BZ), whereas the other two are flat. Each flat band is tangent to one of the dispersive bands at the $\Gamma$ point, as  Fig. \ref{fig:textures} (a) illustrates.

Our results for the orbital Hall conductivities ($\sigma^{z}_{OH}$), calculated as functions of energy by means of the Kubo formula~\cite{mahan}, with the orbital current defined as $J^{\ell^z}_y = \frac{1}{2}\{\ell^z,v_y\}$, are shown in Fig. \ref{fig:textures} (b) for $V_{AB}=0.0$ (blue line), and for $V_{AB} = 0.8$ (red line). Details of these calculations are given in Sec. II of SM. Here we notice a strong orbital Hall conductivity, which peaks at energies close to where the flat bands touch the dispersive bands at $\Gamma$.  For $V_{AB} \ne 0$, the electronic structure develops an energy gap around $E=0$ that eliminates the original Dirac cones in the vicinities of $K$ and $K'$. The flat bands, however, remain tangent to the dispersive bands at $\Gamma$, as shown in  Fig. \ref{fig:textures} (a), and the large OHE in this case also occurs for energies close to where they touch each other. 

The insets of  Fig. \ref{fig:textures} (b) depict the in-plane contribution to the orbital angular momentum textures, calculated on a circle around the $\Gamma$ (left inset) and $K$ (right inset) symmetry points of the 2D first BZ. They are both computed for $V_{AB}=0$. The colours of the arrows emphasise their in-plane azimuthal angles.  At the $\Gamma$ point, the texture displays a dipole-field like structure, whereas in the vicinity of the $K$ points it is identical to the spin-texture  produced by the Dresselhaus SOC in zinc blende lattice systems~\cite{DresslhausNonCentroSymm}. Here, the texture is not caused by SOC, but results only from the orbital features and crystalline symmetry, as we shall subsequently show.

\begin{figure}[h]
	\centering
	\includegraphics[width=0.95\linewidth,clip]{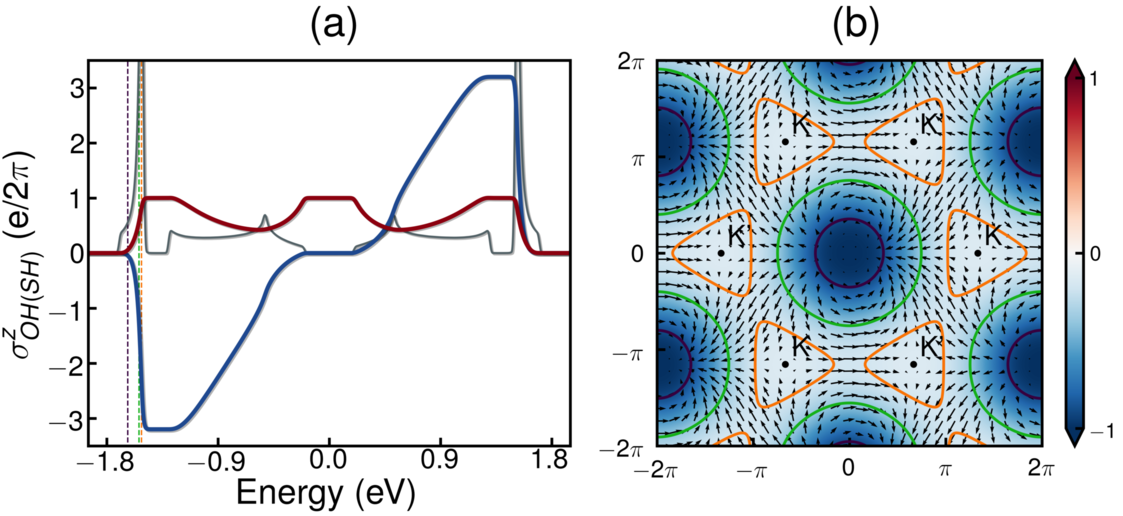}
	\includegraphics[width=0.95\linewidth,clip]{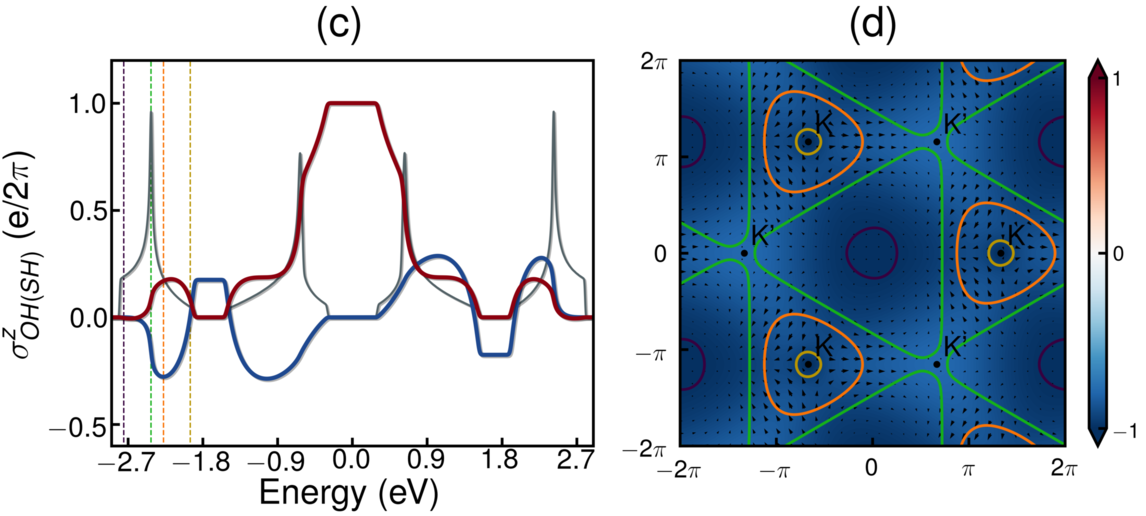}
	\includegraphics[width=0.95\linewidth,clip]{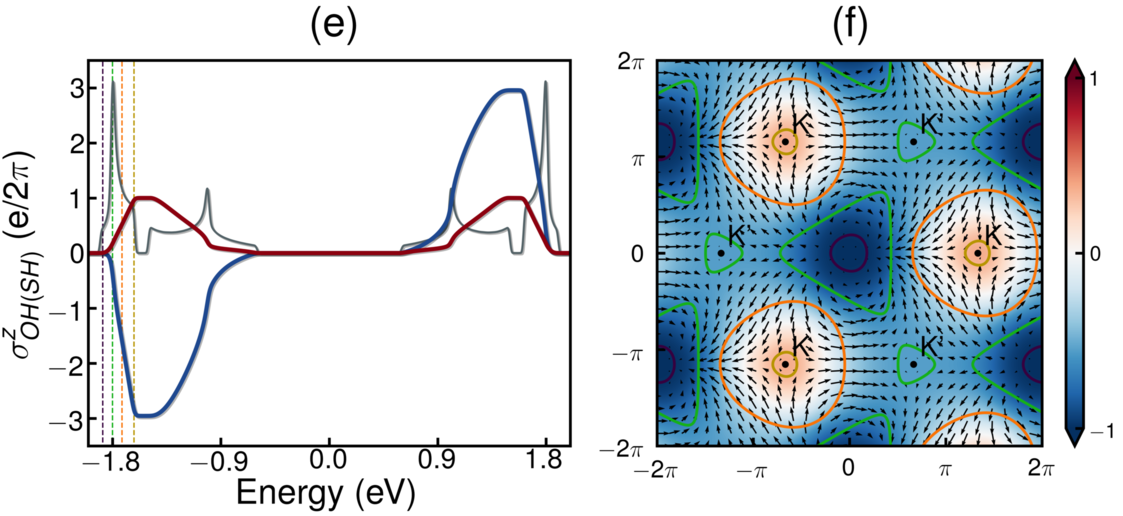}
\caption{Spin Hall conductivity (red), and orbital Hall conductivity (blue), together with the density of states (grey), calculated as functions of energy for: (a)  $\lambda_{I}=0.2$ and $V_{AB}=0$ - B1 phase; (c) $\lambda_{I}=1.1$, and $V_{AB}=0.8$ - B2 phase, and (e) $\lambda_{I}=0.2$, and $V_{AB}=0.8$ - A1 phase. The densities of states are depicted in arbitrary units. Panels (b), (d) and (f)  show the associated orbital textures, calculated for the lower $\uparrow$-spin band, with the same sets of parameters, respectively. The density plots illustrate their corresponding $\langle\ell^z\rangle$ polarisations.}
	\label{fig:soc}
\end{figure}

In the presence of SOC, three energy gaps open: one originating from the $K(K')$ points, and the other two at $\Gamma$, while the flat bands acquire a slight energy dispersion - see  Fig. SI of the SM. When the relative values of $\lambda_{I}$ and $V_{AB}$ vary, this model exhibits a rich topological phase diagram~\cite{MultiorbitalHoneycomb-Wu}. We shall focus on three phases that display distinct topological gap features. They are classified by sets of spin Chern numbers (i,j,k,l) associated with the four $\uparrow$-spin bands, namely  A1 (1,-1,1,-1), B1 (1,0,0,-1), and  B2 (0,1,-1,0), according to the notation of Ref. \onlinecite{MultiorbitalHoneycomb-Wu}. We remind that the spin Chern numbers for the $\downarrow$-spin sector have opposite signs. This codification clearly indicates that when the system is in the A1 phase the two lateral energy gaps are topological, but the central one is not. The reverse occurs in the B2 phase, where only the central energy gap is topological. Last but not least, all the three energy gaps are topological in the B1 phase. This is explicitly verified in the left panels of Fig. \ref{fig:soc}, which show the spin Hall conductivities ($\sigma^z_{SH}$) (red curves) calculated as functions of energy for three different sets of parameters that simulate systems in each of these phases. In the absence of sublattice asymmetry ($V_{AB} = 0$) and for $\lambda_I =0.2$, the system assumes the B1 phase and becomes a QSHI within all the three energy gaps, as the quantised plateaux of $\sigma^z_{SH}$ in Fig. \ref{fig:soc} (a) show. For $\lambda_{I}=1.1$ and $V_{AB}=0.8$, the system is in the B2 phase, which exhibits a quantised spin Hall conductivity plateau in the central energy gap, and two non topological side gaps within which it behaves as an ordinary insulator, displaying no QSHE, as Fig. \ref{fig:soc} (c) illustrates. For $\lambda_{I}=0.2$ and $V_{AB}=0.8$, the system takes on the A1 phase, where it becomes a QSHI for energies within the lateral energy gaps, but behaves as a conventional insulator inside the central gap, as portrayed in  Fig. \ref{fig:soc} (e).

The corresponding orbital Hall conductivities ($\sigma^z_{OH}$) calculated for the three phases (blue curves) are also depicted in the left panels of Fig. \ref{fig:soc}, together with the respective densities of states (grey lines) represented in arbitrary units. We notice that within the lateral gaps, $\sigma^z_{OH}$ exhibits plateaux with much higher intensities than those of the SHE. However, in contrast with the latter, the OHE is not quantised. Its plateaux heights depend upon  $\lambda_I$ and $V_{AB}$, increasing in modulus as the gap width reduces, though limited by the OHE value for $\lambda_I \rightarrow 0$ (see Sec. IV of SM). A remarkable result illustrated in Fig. \ref{fig:soc} (c) is the existence of finite OHE within the two (non-topological) side energy gaps of phase B2, where the system becomes an ordinary insulator with no QSHE. This is particularly interesting because there are no electronic edge states crossing these energy gaps (see Sec. V of the SM), and raises the question on how the orbital Hall current propagates through the system in this case. It is also noticeable that the OHE is an odd function of the Fermi energy ($E_F)$ and vanishes in the central energy gaps for all three phases. This is due to symmetry limitations of this simplified model, which we shall address subsequently.

It is also important to examine how disorder affects the transport properties of these systems. To simulate it we consider on-site potentials $\epsilon_i$ with values randomly distributed within [-W/2, W/2], where W is the disorder strength. We calculate the spin- and orbital-Hall conductivities for different values of W using Chebyshev polynomial expansions and the Kubo-Bastin formula, which are efficiently implemented in the open-source software KITE.  Similarly to what we have previously found for the SHE~\cite{PRLPxPy-Nosotros}, the orbital Hall plateaux are robust to relatively strong Anderson disorder. Details of these calculations are described in sections VI, VII and VIII of the SM.

We know that OHE is linked to orbital textures in reciprocal space~\cite{Go-OHETexture}, and to establish this relationship we have calculated these textures for the $\uparrow$-spin lowest energy band in the entire 2D first BZ. The results are shown in panels (b), (d), and (f) for systems in the B1, B2, and A1 phases, respectively. The orbital characters for all $\uparrow$-spin eigenstates are depicted in Fig. SV of the SM. It is worth noticing that when either $\lambda_I$ or $V_{AB}$ are different from zero, the orbital textures display finite out-of-plane components for each spin direction. However, due to time reversal symmetry the $\ell^z$ orbital polarisations for inverted spin directions are opposite, and consequently the total $\ell^z$ polarisation vanishes. The structure of the in-plane texture, nevertheless, remains the same for both spin components, which means that the in-plane orbital texture survives. It is also noteworthy that both the lowest two energy bands as well as the upper ones display opposite in-plane orbital textures for this simple model. Consequently, the OHE vanishes at the onset of the central energy gap, where the accumulated in-plane orbital texture of the occupied states becomes zero. The absence of electronic states within an energy gap leads to a constant value for $\sigma^z_{OH}$~\cite{Inoue-OHEdTransitionMetal,Streda_QHE,Aires-FermiSeaCorrections} in its range, which justifies the lack of OHE in the central energy gap found for the three phases.

Contour curves are also shown for certain values of  $E_F$ ranging from the bottom of the energy band to the beginning of the lowest energy gap. In all phases, we note that close to the $\Gamma$ point, where the lowest energy band value is minimum, there is virtually no in-plane orbital angular momentum texture, and the OHE is very small. As $E_F$ increases the in-plane orbital texture builds up, assuming a dipole-field like configuration. Eventually, when $E_F$ approaches the onset of the first energy gap, it develops a Dresselhaus-like arrangement near the K and K' points, with opposite signs in each valley. 

In order to uncover the \emph{raison d'etre} of these exotic orbital textures that promote OHE in this systems we derive an effective theory near the Dirac points $K$ and $K'$. Around them, the orbital angular momentum texture is perfectly captured by a linear approximation in the crystalline momentum, whereas it requires a fourth-order expansion near the $\Gamma$ point. Our effective Hamiltonian $\mathcal{H}_{\text{eff}}$ can be expressed in terms of $SU(2) \otimes SU(2)$ orbital and sub-lattice algebras, and written as: $\mathcal{H}_{\text{eff}} = H_{0}+H_{\text{AB}}+H_{\text{SOC}}+H_{\ell}$. Here $H_{0} = -\hbar v_F \left(k_x\sigma_x+\tau k_y\sigma_y\right)$ is the usual Dirac Hamiltonian, with Fermi velocity $v_{F} = \frac{a\sqrt{3}}{2\hbar}$, $a$ denotes the lattice constant, and $\tau = \pm1$ for the $K$ and $K'$ valleys, respectively. $H_{\text{SOC}}=s\lambda_{I}\ell^z$ represents the SOC, where $s=\pm1$ for $\uparrow$ and $\downarrow$ spin electrons, respectively. $H_{\text{AB}}=V_{AB}\sigma^{z}$ is the sub-lattice resolved potential. $H_{{\ell}}$ breaks the degeneracy between $\ell^z$ eigenstates and is given by:

\begin{align}
H_{\ell} = -\frac{\hbar v_{F}}{4}\tau\left( k_{+} \ell_{+}\sigma_{\tau} + k_{-}\ell_{-}\sigma_{\bar{\tau}} \right) -\frac{\sqrt{3}\hbar v_{F}}{2a}\left ( \ell_{x}\sigma_{x}+\tau \ell_{y}\sigma_{y}\right), \nonumber \\
\label{eqn:OrbitalTerm}
\end{align} 
\noindent
where $\sigma_\tau=\sigma_{x}+ i\tau\sigma_{y}$, $\bar{\tau}=-\tau$, $\ell_{\alpha}$ ($\alpha = x,y$) are the orbital angular momentum matrices in the corresponding Hilbert space, $k_{\pm}=k_x\pm i k_y$, and $\ell_{\pm}=\ell_x\pm i \ell_y$.

As shown in the section X of the SM, in the absence of $H_\ell$ each valley presents two degenerated Dirac cones. The first term in the right hand side of Eq. \eqref{eqn:OrbitalTerm} alters the Fermi velocity of the Dirac cones and leads to an in-plane orbital texture profile similar to the one portrayed around the $\Gamma$ point. The second term, however, produces a 
Dresselhaus-like splitting in the Dirac cones and is primarily responsible for the orbital angular momentum texture found in our TB calculations. Our effective theory confirms that the exotic in-plane texture exhibited by these 2D systems is an intrinsic property that arises solely from the $p_x$-$p_y$ orbital characteristics and crystalline symmetries. \begin{figure}[h]
	\centering
	\includegraphics[width=0.99\linewidth]{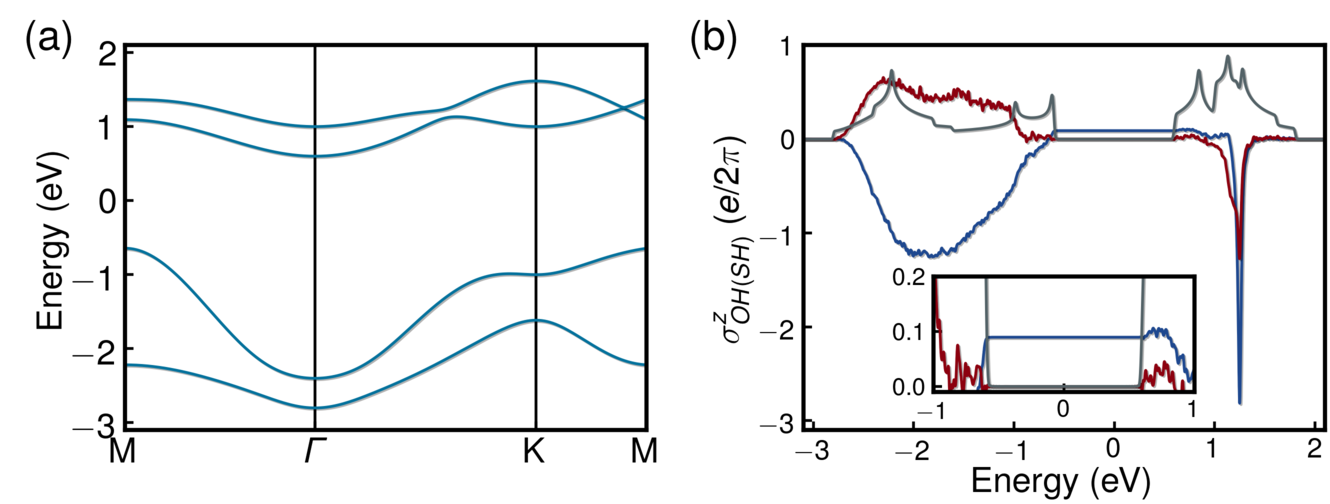}
	\caption{(a) Energy band spectrum calculated for the simple model with second n.n. hopping integrals $V^{(2)}_{pp\sigma}=-0.2$. Here we keep $V_{pp\pi}= 0$, $V_{pp\sigma}= 1$, $\lambda_{I}=0.2$ and $V_{AB}=0.8$. (b) SH (red line) and OH (blue line) conductivities calculated as functions of energy. The grey line depicts the density of states (DOS) in arbitrary units. The inset shows a closeup of the central energy gap highlighting the non-zero value of the OHE within this energy range.}
	\label{fig:tm2}
\end{figure}

\begin{figure}[h]
	\centering
	\includegraphics[width=0.99\linewidth]{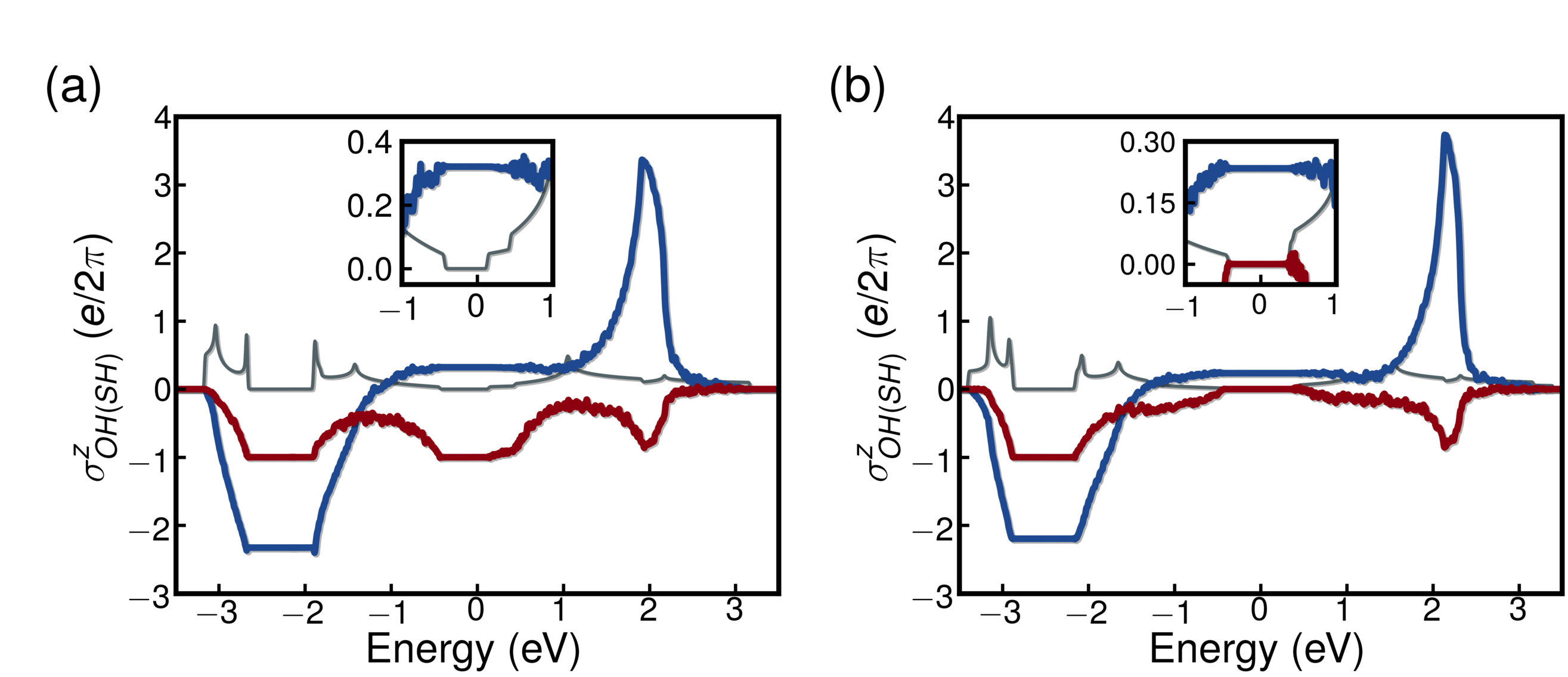}
	\caption{SH (red line) and OH (blue line) conductivities calculated as functions of energies for flat bismuthene: (a) without sublattice asymmetry ($V_{AB}=0$) and (b) with  $V_{AB}=0.87$. The insets highlight the non-zero values of the OHE within the corresponding central energy-gap ranges.The grey line depicts the DOS in arbitrary units.}
	\label{fig:bismt}
\end{figure}

We shall now address the absence of OHE in the central energy gap as results from our calculations. This limitation actually comes from a combination of electron-hole and parity symmetries, which lead to energy levels that are symmetric with respect to the zero energy for this simple model\cite{MultiorbitalHoneycomb-Wu}. One way of breaking it is by introducing second n.n. hopping integrals, as Fig. \ref{fig:tm2} (a) illustrates. Here, just as a proof of concept, we kept $V_{pp\pi}= 0$, and choose the second n.n. hopping integrals $V_{pp\sigma2}=-0.2$. In this case only the central energy gap survives, and within it the system assumes an ordinary insulating phase. Fig. \ref{fig:tm2} (b) clearly shows that the SHE vanishes in this energy range, whereas the OHE is finite. Here, the in-plane orbital texture associated with the second lowest energy band no longer cancels the contribution from the first band. Thus, the OHE does not vanish at the onset of the central energy gap and keeps its non-zero value constant within it. This result, although relatively small in this particular case, unequivocally shows that it is possible to obtain a finite OHE for a non-topological insulating phase, as we previously found for the lateral energy gaps of the B2 phase. Having shown that this effect happens for our simple-model system, it is instructive to inquire into the possibility of observing it in a real system. A candidate is the recently synthesised flat bismuthene grown on SiC, whose low energy electronic properties are reasonably well described by an effective TB model Hamiltonian that includes only two orbitals ($p_x$ and $p_y$) per atom ~\cite{ReisBismutheneExperimental,li2018new,zhou2018giant,antimonene}. It is a real solid state system, typical of a promising class of 2D materials based on the group group VA elements that exhibit relatively large energy gaps. In fact, a very good TB fit of both the valence and conduction bands of flat bismuthene can be obtained with the inclusion of second n.n. hopping integrals, as Fig. SX of the SM illustrates. Results for the associated SHE and OHE calculated as functions of $E_F$ for planar bismuthene employing a Chebyshev polynomial expansion method are shown in Fig.\ref{fig:bismt}. We clearly see in this case that the spectra are not symmetric with respect to the zero energy and the right-hand side gap disappear. Results for $V_{AB} = 0$, depicted in Fig.\ref{fig:bismt} (a), show that the remaining gaps are topological, displaying a quantised SHE, and significant OHE. For sufficiently large sublattice asymmetry, however, the central gap ceases to be topological, exhibiting no SHE, as Fig.\ref{fig:bismt} (b) illustrates. Notwithstanding, the orbital Hall conductivity is appreciable within this energy range. This validates our original prediction that pure orbital angular momentum currents can be triggered by a longitudinally applied electric field in some normal insulators. 

In summary, we have performed detailed analyses of the spin and orbital Hall conductivities for a class of 2D systems, relating the corresponding OHE, SHE and orbital textures. Our calculations show the existence of OHE in topological insulators, with values that exceed those obtained for the SHE. Remarkably, we also obtain OHE for normal insulating phases where the SHE is absent and no edge states cross their energy gaps. We show that the OHE in these systems is associated with exotic momentum-space orbital textures that are caused by an intrinsic Dresselhaus-type of interaction. This is rather general and show that certain 2D insulating materials can generate orbital angular momentum currents that may be useful for developing novel spin-orbitronic devices. 
 
\begin{acknowledgments}
	We acknowledge CNPq/Brazil, FAPERJ/Brazil and INCT Nanocarbono for financial support, and NACAD/UFRJ for providing high-performance computing facilities.  TGR acknowledges COMPETE2020, PORTUGAL2020, FEDER and the Portuguese Foundation for Science and Technology (FCT) through project POCI-01- 0145-FEDER-028114. TPC acknowledges S\~ao Paulo Research Foundation (FAPESP) grant 2019/17345-7. 
	
\end{acknowledgments}

\clearpage

\begin{widetext}
	\begin{center}
		\textbf{\large Supplementary material for ``Spin and Charge Transport of Multi-Orbital Quantum Spin Hall Insulators''}
	\end{center}

\setcounter{equation}{0}
\setcounter{figure}{0}
\setcounter{section}{0}
\setcounter{table}{0}
\setcounter{page}{1}
\renewcommand{\theequation}{S\arabic{equation}}
\renewcommand{\thetable}{S\Roman{table}}
\renewcommand{\thefigure}{S\Roman{figure}}

\section{Hamiltonian in reciprocal space and orbital angular momentum operators definition }
The Hamiltonian given by Eq. (1) of the main text can be rewritten in the reciprocal space. Using the basis $\{\big|p_+,A\big>, \big|p_-,A\big>, \big|p_+,B\big>, \big|p_-,B\big>\} $, the hopping term reads,

\begin{eqnarray}
\tilde{H}_0 (\vec{k})=\left( \begin{array}{cccc}
0 & 0 & \tilde{A}(\vec{k}) & \tilde{D}(\vec{k})  \\
0 & 0 & \tilde{C}(\vec{k}) & \tilde{B}(\vec{k})  \\
\tilde{A}^*(\vec{k}) & \tilde{C}^*(\vec{k}) & 0 & 0  \\
\tilde{D}^*(\vec{k}) & \tilde{B}^*(\vec{k}) & 0 & 0    \end{array} \right) ,\label{UHhopping}  
\end{eqnarray}
where
\begin{eqnarray}
2\tilde{A}(\vec{k})=t_{xx}(\vec{k})+t_{yy}(\vec{k})+i\Big( t_{xy}(\vec{k})-t_{yx}(\vec{k}) \Big), \\
2\tilde{B}(\vec{k})=t_{xx}(\vec{k})+t_{yy}(\vec{k})+i\Big( t_{yx}(\vec{k})-t_{xy}(\vec{k}) \Big), \\
2\tilde{C}(\vec{k})=t_{xx}(\vec{k})-t_{yy}(\vec{k})+i\Big( t_{xy}(\vec{k})+t_{yx}(\vec{k}) \Big), \\
2\tilde{D}(\vec{k})=t_{xx}(\vec{k})-t_{yy}(\vec{k})-i\Big( t_{xy}(\vec{k})+t_{yx}(\vec{k}) \Big),
\end{eqnarray}
The transfer integrals $t_{\alpha\beta}(\vec{k})$ are Fourier transforms of Slater-Koster coefficients in the honeycomb lattice,
\begin{eqnarray}
t_{\alpha,\beta}(\vec{k})=\sum_{m=0}^2 \big< p_\alpha, A\big| \hat{V}\big| p_\beta,B\big>(m) e^{-i \vec{k}\cdot \vec{a}_m}, \label{tfunctions}
\end{eqnarray}
where, $m$ runs over the three nearest-neighbours of a site in sublattice $A$, that are located in sublattice B, and $\vec{a}_m$ is the vector connecting the atom at sublattice A with its mth neighbor at sublattice B. The Slater-Koster integrals are given by
\begin{eqnarray}
t_{\alpha \alpha}(\vec{k})= n_{\alpha}^2(m) V_{pp\sigma}+(1-n_{\alpha}^2(m))V_{pp\pi}, \label{skii} \\
t_{\alpha \beta}=-n_{\alpha}(m)n_{\beta}(m)\big( V_{pp\pi}-V_{pp\sigma} \big),
\end{eqnarray}
with $n_{\alpha,\beta}(m)$ being the direction cosine connecting the site of sublattice $A$ with $m$-th first neighbor on sublattice $B$. In all results presented in this work, we set $V_{pp\pi}=0$, unless it is mentioned. This condition can be relaxed without changing any of the main conclusions of our work. In this basis, the SOC term is diagonal, $H_{soc}^{s}=s\lambda_I \text{diag} (1,-1,1,-1)$ and the sublattice potential is given by  $H_{AB}=V_{AB} \text{diag} (1,1,-1,-1)$.\\

As mentioned on the main text, due the absence of the $p_z$ orbital, the electronic states are restricted to the subspace associated with $m_{\ell}=\pm1$ only, hence the angular momentum operators can be redefined in terms of a $SU(2)-$algebra as:

\begin{align}
l_z=\big|p_+\big>\big<p_+\big|-\big|p_-\big>\big<p_-\big|,\nonumber \\
l_x=\big|p_+\big>\big<p_-\big|+\big|p_-\big>\big<p_+\big|, \nonumber\\
l_y=i\Big(\big|p_-\big>\big<p_+\big|-\big|p_+\big>\big<p_-\big| \Big)
\end{align}

\section{Kubo formula for linear response conductivity }
In the main text, we compute the spin and orbital Hall conductivity for different topological phases o the  Hamiltonian of Eq. (1) in the main text. For the cases of the pristine system, we used Kubo formalism to compute both OHE and SHE. In this formalism, the spin Hall (SH) and orbital Hall (OH) $\eta$-polarized response, in $\hat{y}$ direction, to an electric field applied in the $\hat{x}$ direction is given by,

\begin{eqnarray}
\sigma^{\eta}_{OH(SH)}=\frac{e}{\hbar} \sum_{n\neq m}\sum_{s=\uparrow,\downarrow} \int_{B.Z.} \frac{d^2k}{(2\pi)^2} (f_{m\vec{k}}-f_{n\vec{k}}) \Omega_{n,m,\vec{k},s}^{X_{\eta}}, \label{Kubo1}
\end{eqnarray}
\begin{eqnarray}
\Omega_{n,m,\vec{k},s}^{X_{\eta}}=\hbar^2 \text{Im} \Bigg[ \frac{\big<\psi^s_{n,\vec{k}}\big|j_{y}^{X_{\eta}}(\vec{k})\big|\psi^s_{m,\vec{k}}\big>\big<\psi^s_{m,\vec{k}}\big|v_x(\vec{k})\big|\psi^s_{n,\vec{k}}\big>}{(E^s_{n,\vec{k}}-E^s_{m,\vec{k}}+i0^+)^2}\Bigg] \label{Kubo2}
\end{eqnarray}

Were $\sigma^{\eta}_{OH(SH)}$ is the orbital Hall (Spin Hall) DC conductivity with polarization in $\eta$-direction, $\Omega_{n,m,\vec{k},s}^{X_{\eta}}$ is the related gauge-invariant Berry curvature. In Eq. \ref{Kubo2}, $E^s_{n(m),\vec{k}}$ and $|\psi^s_{n(m),\vec{k}}\big>$ are eigenvalues and eigenvectors of Hamiltonian of Eq. (\ref{UHhopping}), for $n(m)$ Bloch band, with $n,m=1,..,4$ (in crescent order of energy), and $s=\uparrow,\downarrow$ spin-sector. Velocity operators are defined by, $v_{x(y)}(\vec{k})=\partial H(\vec{k})/\partial k_{x(y)}$, where $H(\vec{k})$ is the tight-binding Hamiltonian in reciprocal space. The current operator in $\hat{y}$ direction is defined by $j_y^{X_{\eta}}(\vec{k})=\big(X_{\eta}v_y(\vec{k})+v_y(\vec{k})X_{\eta} \big)/2$, where $X_{\eta}=\hat{\ell}_{\eta}(\hat{s}_{\eta})$ for OH (SH) conductivities polarized in $\eta$ direction.

As it was discussed in the main text, this model presents a non-vanishing $\sigma^{z}_{OH}$, even in absence of SOC, in contrast to the spin Hall ($\sigma^{z}_{SH}$) response which depends on the presence of SOC or exchange interaction. Added to this, in the presence of an exchange term, it was shown that the  model presents a non-vanishing $\sigma^{x}_{OH}$ associated with in-plane polarized orbital Hall effect. It is important to mention that equations \ref{Kubo1} and \ref{Kubo2} are valid only in the clean limit and do not take into account the effect of disorder. However, as was briefly pointed in the main text, the effect of disorder should not affect our results for insulating phases (Fermi energy inside an electronic gap) due to the absence of the Fermi surface, responsible to generates the leading-order contribution in the computation of vertex corrections \cite{Aires-FermiSeaCorrections}. We confirm the robustness of our results against Anderson disorder using a real-space computation method which is discussed in the next sections of the SM.

\section{Analysis of the band structure}

We have examined the orbital Hall conductivity properties of three distinct topological phases displayed by the Hamiltonian $H$ defined by Eq. (1) in the main text. They are labelled as B1, A1, and B2 phases, according to the classifications used in Ref.~\onlinecite{MultiorbitalHoneycomb-Wu}. Figure \ref{fig:bands} shows the  $\uparrow$-spin electron energy bands for the system in these three phases. The spin-$\downarrow$ bands can be deduced by applying a time-reversal symmetry operation on $H$. Panel (a) illustrates the band structure of the B1 phase, calculated for $\lambda_{I} = 0.2V_{pp\sigma}$, and $V_{AB}=0$. We notice that the SOC causes three energy gaps to open, one originating from the $K$($K'$) points, and the other two at $\Gamma$, while the flat bands acquire a slight energy dispersion. Panel (b) shows the energy bands for the system in the A1 phase, calculated with $\lambda_{I} = 0.2V_{pp\sigma}$, and $V_{AB}=0.8V_{pp\sigma}$. The sub-lattice potential affects each valley differently, as expected, because it breaks the degeneracy between eigenvalues at the $K$ and $K'$ symmetry points. By examining the opposite spin polarisation one finds that this phase exhibits a strong spin-valley locking, as discussed in Refs.~ \onlinecite{MultiorbitalHoneycomb-Wu,zhou2018giant,PRLPxPy-Nosotros}. Panel (c) displays the energy bands for the system in the B2 phase, calculated with $\lambda_{I}=1.1V_{pp\sigma}$ and $V_{AB}=0.8V_{pp\sigma}$. In this case, $\lambda_{I}$ is comparable but slightly larger than $V_{AB}$, and we note that they lead to effects that are similar to those exhibited panel (b), including a strong spin-valley locking with valley polarisation stronger than in the previous case due to the relatively large values of $\lambda_{I}$ and $V_{AB}$.

\begin{figure}[h]
	\includegraphics[width=0.85 \columnwidth]{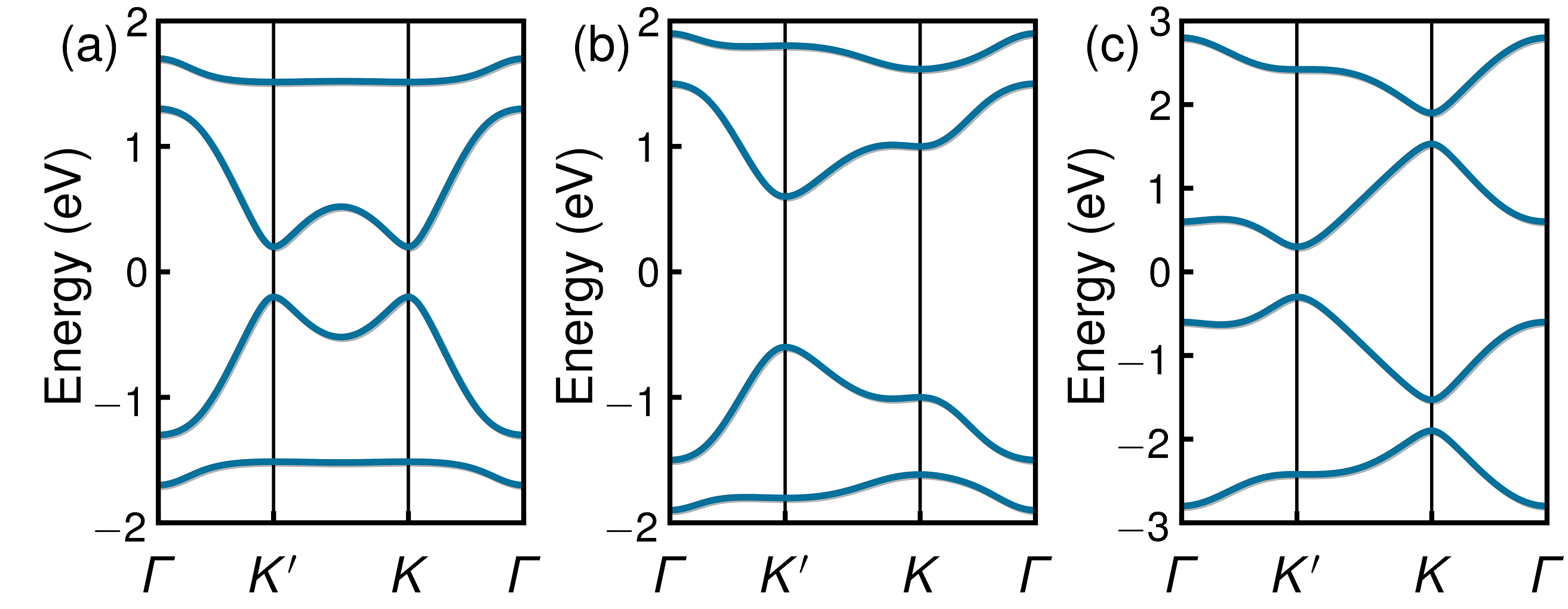}
	\caption{$\uparrow$-spin electron energy bands calculated as functions of wave vectors along some symmetry directions in the 2D Brillouin zone for three distinct topological phases: (a) B1 with $\lambda_{I}=0.2V_{pp\sigma}$ and $V_{AB}=0$. (b) A1 with $\lambda_{I}=0.2V_{pp\sigma}$ and $V_{AB}=0.8V_{pp\sigma}$ (c) B2 with $\lambda_{I}=1.1V_{pp\sigma}$ and $V_{AB}=0.8V_{pp\sigma}$.}
	\label{fig:bands}
\end{figure}

\section{Evolution of the Orbital Hall effect plateaux}
\begin{figure}[ht]
	\centering
	\includegraphics[width=0.85\columnwidth]{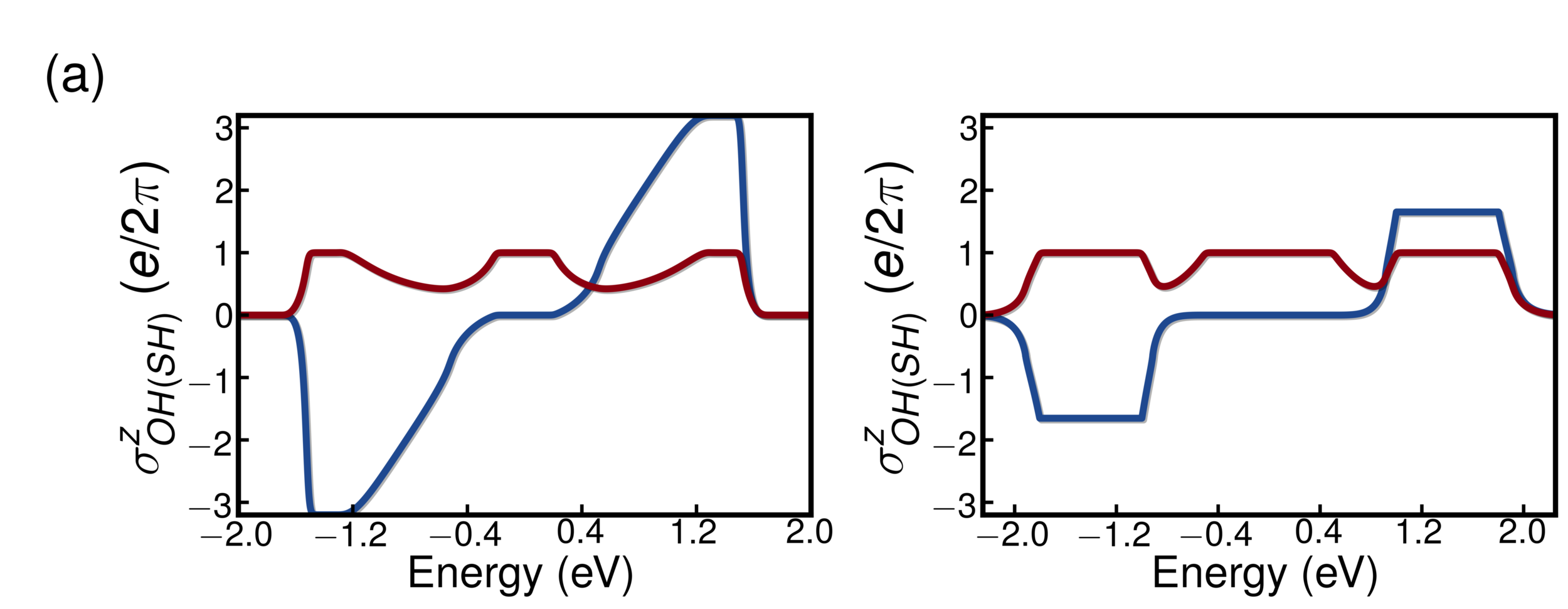}
	\includegraphics[width=0.85\columnwidth]{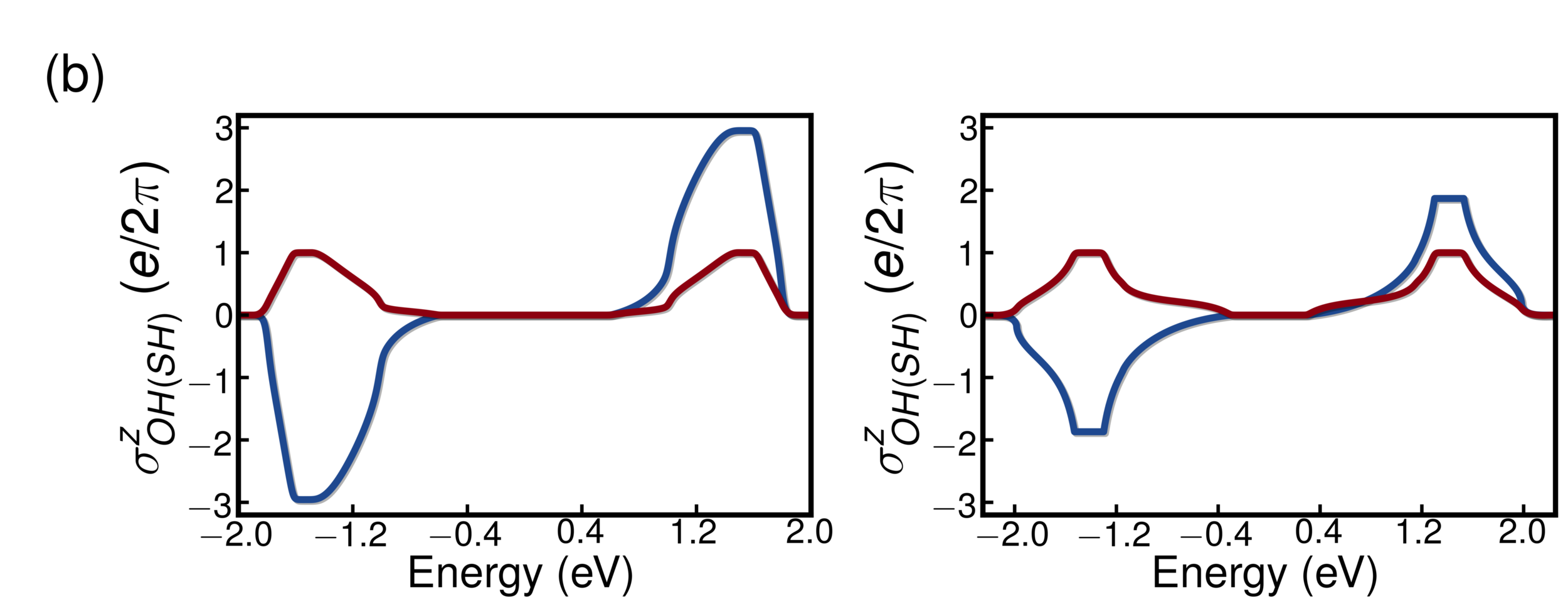}
	\includegraphics[width=0.85\columnwidth]{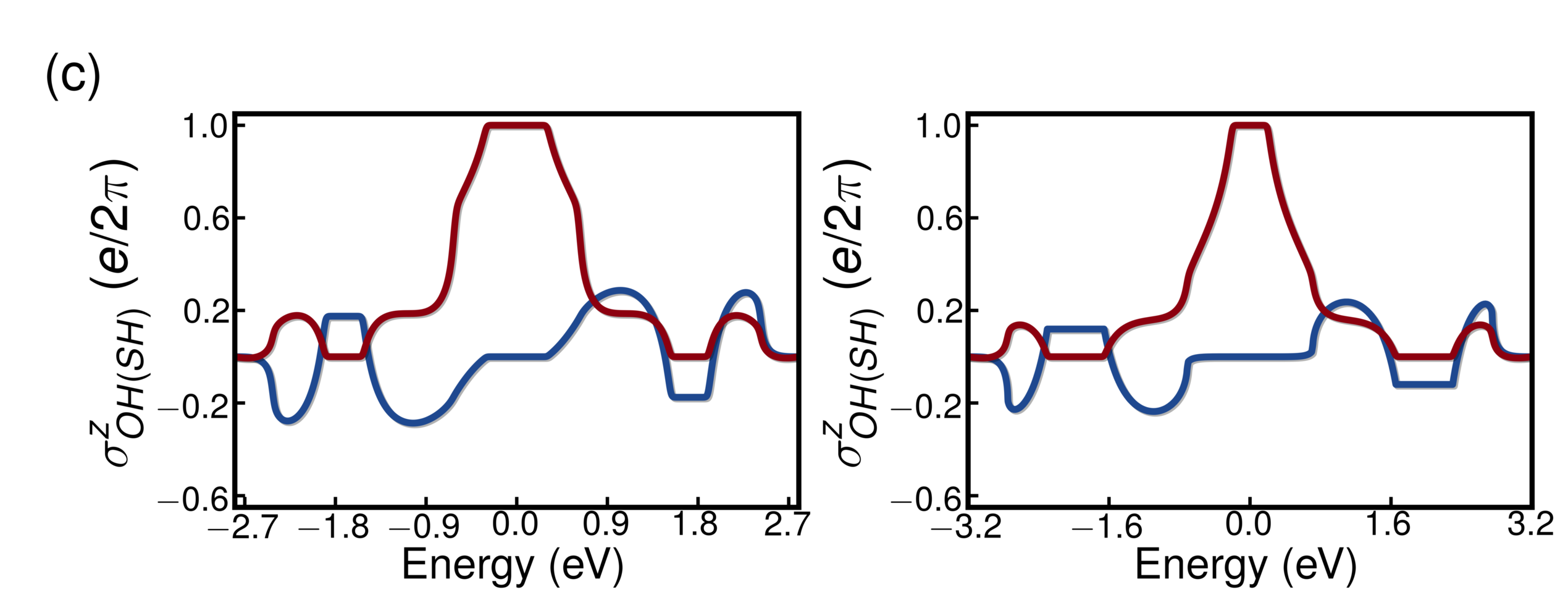}
	\centering
	\caption{ Spin Hall conductivity $\sigma^z_{SH}$ (red) and orbital Hall conductivity $\sigma^z_{OH} (blue)$ calculated for: (a) $V_{AB}=0$ and $\lambda_{I}=0.2V_{pp\sigma}$ (left) and $\lambda_{I}=1.0V_{pp\sigma}$ (right). (b) $V_{AB}=0.8V_{pp\sigma}$ and $\lambda_{I}=0.2V_{pp\sigma}$(left)  and $\lambda_{I}=0.5V_{pp\sigma}$ (right). (c)  $V_{AB}=0.8V_{pp\sigma}$,  $\lambda_{I}=1.1V_{pp\sigma}$ (left) and $\lambda_{I}=1.5V_{pp\sigma}$ (right)}	
	
	\label{fig:plateau}
\end{figure}
In the main text, it was mentioned that the height of the orbital Hall plateaux within the lateral gaps depends upon the SOC coupling constant and the sub-lattice resolved potential. To demonstrate this, we show in Figure \ref{fig:plateau} results for the spin and orbital Hall conductivities calculated for different sets of parameters for the B1, A1, and B2 phases. The results depicted in each panel of Figure \ref{fig:plateau} are obtained for a fixed value of $V_{AB}$ and two different values of $\lambda_{I}$ that are represented in the left and right columns, respectively. In panel (a) we show the conductivities calculated for $V_{AB}$=0$,\lambda_{I}=0.2V_{pp\sigma}$ and $\lambda_{I}=1.0V_{pp\sigma}$, which correspond to situations in which the system is in the B1 phase. It is clear that the height of the OHE plateau decreases as the SOC increases. In fact, the height of the plateau scales with the size of the lateral gap, being close to the maximum value of the metallic limit for very small gaps. The same trend is observed in the other two phases, in contrast with the heights of the spin Hall plateaux that remain the same in all cases .

\section{Zigzag Nano-ribbons spectra}

\begin{figure}[h]
	\includegraphics[width=0.9\columnwidth]{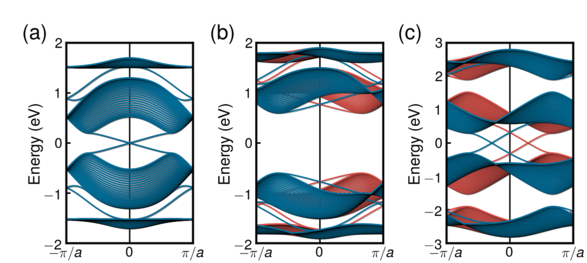}
	\caption{Zigzag Nano-ribbons spectra for phases (a) $B1$ with $\lambda_{I}=0.2V_{pp\sigma}$ and $V_{AB}=0$, (b) A1 with $\lambda_{I}=0.2V_{pp\sigma}$ and $V_{AB}=0.8V_{pp\sigma}$, and (c) B2 with $\lambda_{I}=1.1V_{pp\sigma}$ and $V_{AB}=0.8V_{pp\sigma}$.}
	\label{fig:ribbon}
\end{figure}

Thorough the main text we analysed the spin and orbital Hall effects for the three distinct topological phases $B1,A1\text{ and } B2$. To further substantiate our findings of the non-zero orbital Hall conductivity in the trivially insulating phases, we analysed the energy spectrum of a zigzag nano-ribbon in our system for the three distinct phases. Figure \ref{fig:ribbon} shows the spectra for each of the phases studied in the main text. As expected,  the number of pairs of edge states corresponds with the index $\mathbb{Z}_{2}$ of each of these phases. Panel (a) shows the energy bands corresponding to the phase $B1$, here the most interesting features are the pairs of edge channels that cross the gap and the fully symmetric spectrum for both spin polarizations. Panel (b) displays the spectrum of the $A1$ phase, here we can see the strong spin-valley locking that results from the inversion symmetry breaking produced by staggered sub-lattice potential. Interestingly here we can observe the absence of edge states traversing the central gap. Finally panel (c) shows the band structure of a ribbon in the phase $B2$. Here we can see how due the strong spin-orbit coupling and staggered sub-lattice potential the edge states in both the lateral gaps do not cross the gap while the edge estates of the central gap are crossing again. The results are fully consistent with the spin Cher number characterization of these phases done in Ref~\cite{MultiorbitalHoneycomb-Wu}. The results of panel (c) are the most striking ones, because they indicate that differently from the spin Hall conductivity, the orbital Hall effect plateau does not require electronic conducting channels to have a constant non-quantized value.

latex onecolumn undefined

\section{Chebyshev Polynomial Expansion}

To study the transport and spectral properties of the honeycomb lattice with $p_x-p_y$ orbitals we used the Chebyshev polynomial expansion. In this numerical method, the Green and spectral functions are accurately expanded in terms of Chebyshev polynomial of first kind of the Hamiltonian matrices\cite{SilverKPM,WeisseKPM}. This set of polynomials are commonly chosen due their unique convergence properties, their relation with the Fourier transform and their convenient recurrence relations that allows the iterative construction of higher order polynomials\cite{SilverKPM,WeisseKPM}. In  recent years this method has gained much attention in the study of the transport properties of 2D systems\cite{Aires-JoaoKPM,KPM-DisorderGraphene,Graphene-Aires-Largescale, JoseTMDGraphene, Canonico2018}.  Because of  its high scalability, It was used to study of topological phase transitions induced by disorder\cite{PRLPxPy-Nosotros}, and more recently, in the analysis of the electronic properties of  graphene encapsulated between two twisted hBN structures\cite{andjelkovic2019double}. The method requires a rescaling of the Hamiltonian and it spectrum to make them fit into the interval $(-1,1)$ where the Chebyshev polynomials are defined and consequently the convergence of the method is assured. This scaling is achieved by means of the transformations $\tilde{\mathcal{H}} = \left(\mathcal{H}-b\right)/a$ and $\tilde{E} = \left(E-b\right)/a$ where $a\equiv\left(E_{T}-E_{B}\right)/\left(2-\epsilon\right)$ and $b\equiv\left(E_{T}+E_{B}\right)/2$. In the later $E_{T}$ and $E_{B}$ represents the top and bottom limits of the spectrum, respectively, and $\epsilon$ is a small cut-off parameter introduced to avoid numerical instabilities.\\

With this later conditions fulfilled, the Chebyshev polynomial expansion of the density operator considering $N$ polynomials can be written as:

\begin{equation}
\rho\left(\tilde{E}\right) = \frac{1}{\pi\sqrt{1-\tilde{E}^2}}\sum_{m=0}^{N-1}g_{m}\mu_{m}T_{m}\left(\tilde{E}\right),
\label{eqn:densityopcheb}
\end{equation}

where $g_{m}$ is a kernel introduced to control the Gibbs oscillations produced by the sudden truncation of the series expansion\cite{SilverKPM,WeisseKPM}. The coefficients are calculated with $\displaystyle \mu_{m} = \langle Tr T_{m}\left(\tilde{H}\right)\rangle$, in which $\langle\dots \rangle$ represents the average over different disorder configurations. The calculation of the density operator of a given system is reduced to the computation of the trace of a matrix. To further decrease the computational cost of the calculation of quantities such as the density operator, instead of calculating the full trace of the polynomial matrices\cite{WeisseKPM}, we simply approximate the expansion coefficient $\mu_{m}$ as

\begin{equation}
\mu_{m} \approx \frac{1}{R}\langle \sum_{r=1}^{R}\langle\phi_{r}|T_{m}\left(\tilde{\tilde{\mathcal{H}}}\right)|\phi_{r}\rangle \rangle
\end{equation}

In the later $|\phi_{r}\rangle$ represent a set of random vectors which are defined as $|\phi_{r} \rangle = D^{-1/2}\sum_{i=1}^{D} e^{i\phi_{i}}|i \rangle$. Here $\left\lbrace|i\rangle\right\rbrace_{i=1,\dots,D}$ denotes the original basis set, in which orbitals and spins on the lattice sites are treated equivalently, $D$ represents the dimension of the Hamiltonian matrix, and $\phi_{i}$ is the phase of each of the state vectors that comprise each of the random vectors. $R$ is the number of random vectors used in the trace estimation and the convergence of the later goes as $1/\sqrt{DR}$.

\section{Chebyshev polynomial expansion of Kubo Formula}

To compute the spin and orbital conductivities of disordered systems, we employed the efficient algorithm developed by J. Garc\'ia et. al.\cite{Jose-TatianaKPM,Jose-tatiana}, which is based in the Chebyshev expansion of the Kubo-Bastin formula\cite{Bastin-FormulaConductividad}:

\begin{flalign}
&\sigma_{\alpha \beta}(\mu, T) = \frac{i\hbar}{\Omega}\int_{-\infty}^{+\infty}dE f(E; \mu, T) \nonumber\\ &\times Tr \langle j_{\alpha}\delta(E-\mathcal{H})j_{\beta}\frac{dG^{+}}{dE} - j_{\alpha}\frac{dG^{-}}{dE}j_{\beta}\delta(E-\mathcal{H})\rangle, 
\label{eqn:	KB}
\end{flalign}

in which $\Omega$ represents the area of the $2D$ sample, $f(E; \mu, T)$ is the Fermi-Dirac distribution for the energy $E$, chemical potential $\mu$ and temperature $T$. $G^{+}(G^{-})$ symbolise the advanced(retarded) one electron Green function. As it can be seen from \eqref{eqn:	KB} the Kubo-Bastin formula is expressed as a current-current correlation function. Then, to adapt this formula to calculate the spin hall conductivity $\sigma_{SH}^z $, we define $j_{\alpha}$ as the current-density operator like $j_{\alpha} \equiv j_{x} = \frac{ie}{\hbar}\left[x,\mathcal{H}\right]$ and $j_{\beta}$ as the spin current-density as $j_\beta \equiv j_{y}^s = \frac{1}{2}\left\lbrace\sigma_{z},v_{y}\right\rbrace$ where $\sigma_{z}$ is the usual Pauli's matrix and $v_{y}$ is the $y$-Component of the velocity operator. For the computation of the orbital Hall conductivity, again $\sigma_{OH}^z$ we define the current operator $j_{\alpha}$ as  $j_{\alpha} \equiv j_{x} = \frac{ie}{\hbar}\left[x,\mathcal{H}\right]$ and we write $j_{\beta}$ as the orbital current density operator, which is defined like $j_\beta \equiv j_{y}^s = \frac{1}{2}\left\lbrace \ell_{
	z},v_{y}\right\rbrace$ where $\ell_{z}$ is the $z$. It is noteworthy to mention that for the spin hall conductivity calculations we used the open-source code from the KITE project\cite{KITE}.

\section{Numerical simulation of the disordered case }

It is instructive to investigate how disorder affects the OHE in these two-dimensional systems and more specifically, how it modifies the plateaux in the orbital Hall conductivity that, as discussed before, is not dominated by conducting edge states. For this purpose we include in our Hamiltonian an on-site Anderson disorder term $\epsilon_{i}$ whose values are randomly picked from an uniform distribution that goes from $\left[-\frac{W}{2},\frac{W}{2}\right]$, in which $W$ represents the Anderson disorder strength and then proceeded with the aforementioned Chebyshev polynomial expansions to compute the density of states (DOS), and the transverse components of the spin and orbital conductivity tensors. In these calculations we have considered systems of $8\times 256 \times 256$ orbitals, Chebyshev polynomials up to the order $M=1280$ and we averaged over $R=150$ random vectors. It is noteworthy to mention that due the large number lattice sites that we are considering, we restricted ourselves to only one disorder realization, this is based on the assumption that almost every possible configuration is contained on our system due it large size.

\begin{figure}[h]
	\centering
	\includegraphics[width=0.9\columnwidth]{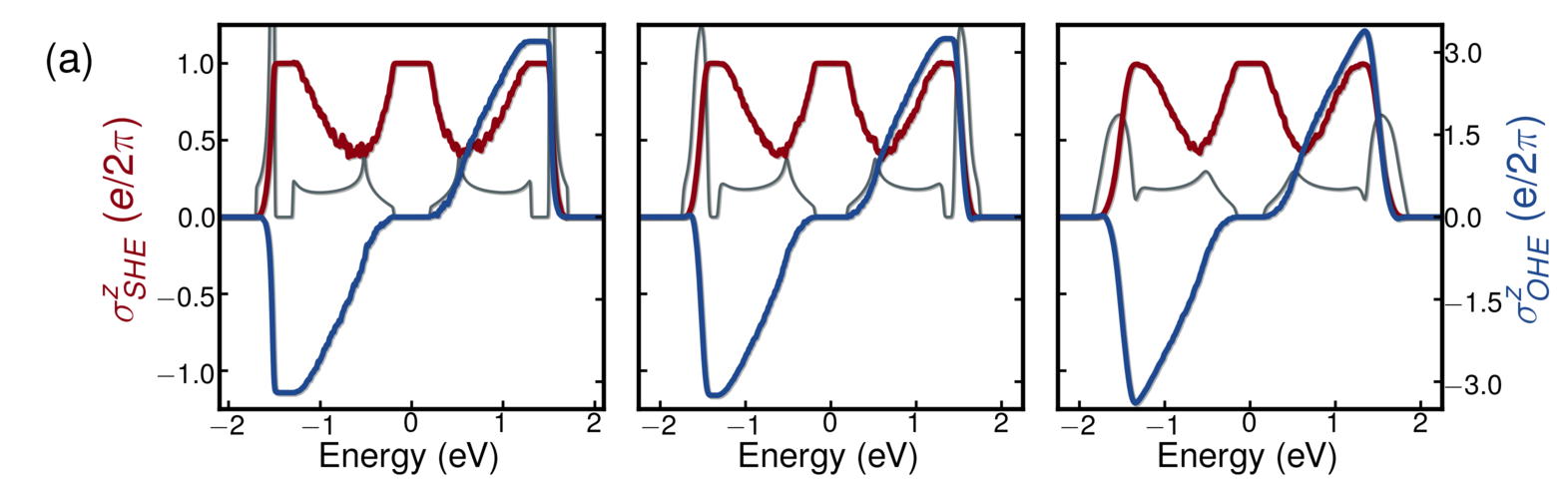}
	\includegraphics[width=0.9\columnwidth]{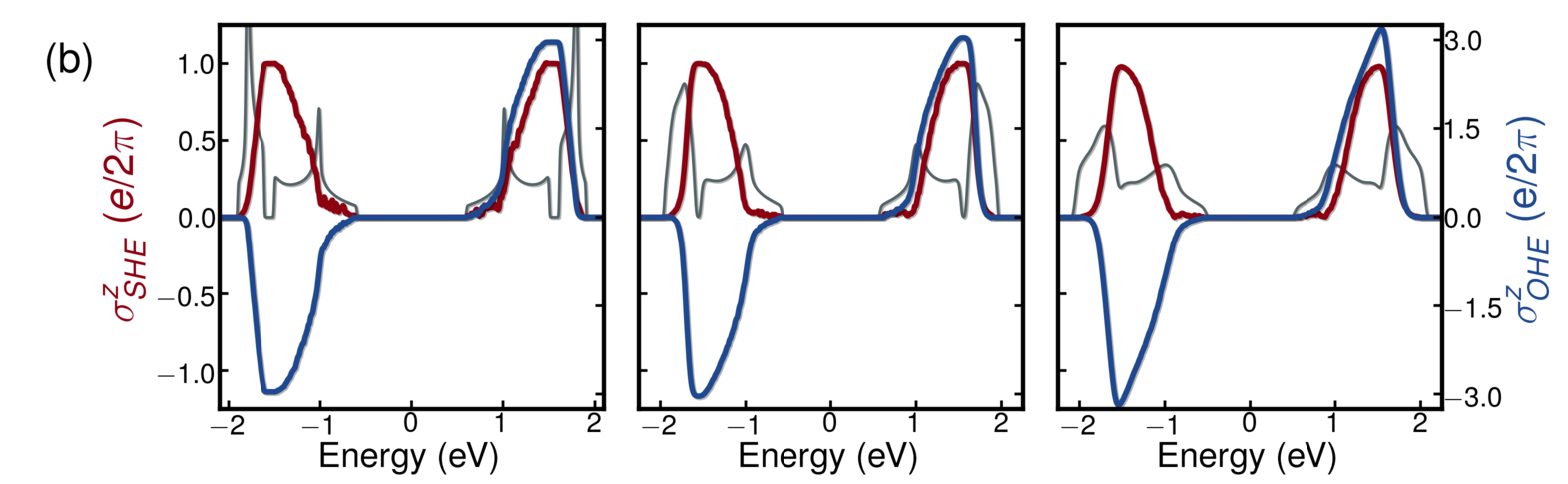}
	\includegraphics[width=0.9\columnwidth]{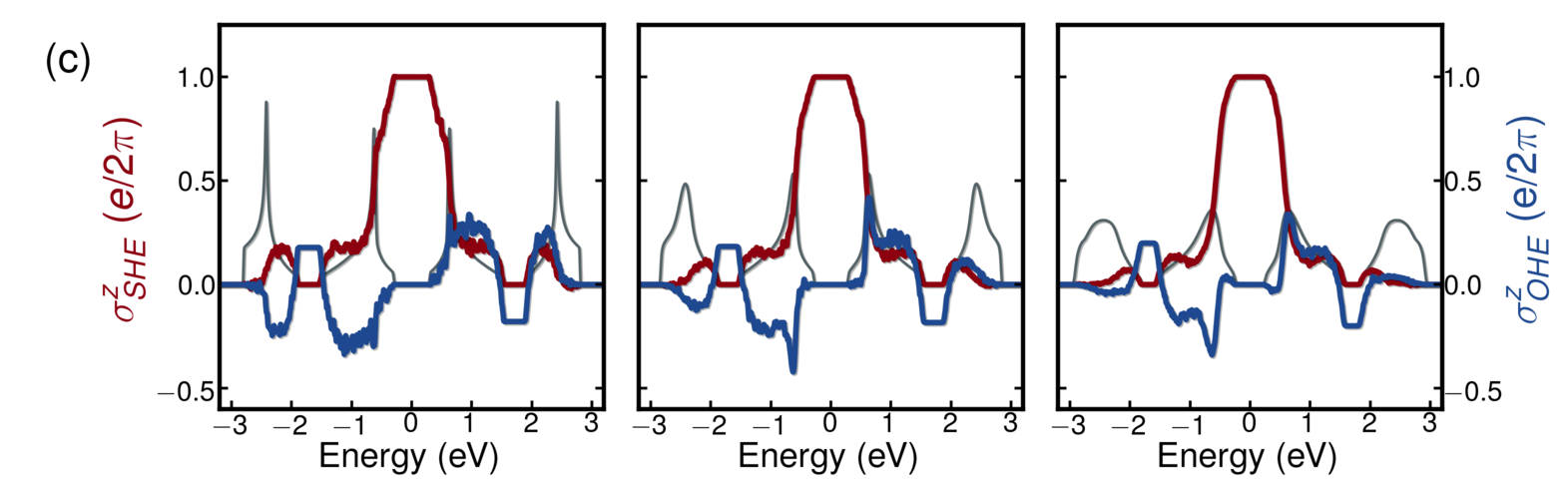}
	\caption{Spin (red) and orbital (blue) Hall conductivities calculated as functions of energy for: (a) $\lambda_{I}=0.2V_{pp\sigma}$ $V_{AB}=0$, (b) $\lambda_{I}=0.2V_{pp\sigma}$ $V_{AB}=0.8V_{pp\sigma}$, and (c) $\lambda_{I}=1.1V_{pp\sigma}$ $V_{AB}=0.8V_{pp\sigma}$ in the presence of disorder. The left, central and right panels show the results obtained in the relatively weak ($W=0.05V_{pp\sigma}$), intermediate ($W=0.2V_{pp\sigma}$) and strong ($W=0.4V_{pp\sigma}$) disorder regimes, respectively. The grey lines represent the density of states calculated for the same set of parameters. \label{fig:disorder}}
\end{figure}

\pagebreak

Figure  \ref{fig:disorder} shows the spin and orbital Hall conductivities calculated for both weak ($W=0.05V_{pp\sigma}$), intermediate ($W=0.2V_{pp\sigma}$) and strong ($W=0.4V_{pp\sigma}$) disorder. Similarly to what was previously observed for the SHE~\cite{Canonico2018}, the orbital Hall plateau remains present, even for a relatively strong disorder that closes the lateral gaps. Our preliminary results indicate that the orbital Hall effect in two-dimensional insulators is robust against Anderson disorder.

\section{Orbital Texture Analysis}
In contrast with the SHE, our calculations show that the OHE is not quantised, and occurs even in the absence of metallic edge states. In order to explore the origin of the OHE in this model system, we investigated the characteristics of its orbital angular momentum in reciprocal space within the 2D first BZ. To this end, we compute the orbital texture in reciprocal space defined as,
\begin{eqnarray}
\vec{L}^s_{n,\vec{k}}=\big<\ell_x\big>_{n,\vec{k}}^s\hat{x}+\big<\ell_y\big>_{n,\vec{k}}^s\hat{y}+\big<\ell_z\big>_{n,\vec{k}}^s\hat{z},
\end{eqnarray}
Where, $\big<\ell_{x,y,z}\big>_{n,\vec{k}}^s=\big<\psi^s_{n,\vec{k}}\big|\ell_{x,y,z}\big|\psi^s_{n,\vec{k}}\big>$ is the expected value of angular-momentum operator in reciprocal space for states of Bloch band $n$ and spin sector $s$. To study the orbital texture and how it affects the OHE, we separate the in-plane textures ($\big<\ell_{x,y}\big>_{n,\vec{k}}^s$), which are represented by arrows, see Fig. \ref{fig:SMtextures} and panels of Fig.2 of main text, and out-of-plane textures ($\big<\ell_{z}\big>_{n,\vec{k}}^s$), which we represent as a color plot (dark blue color for $\big<\ell_{z}\big>\approx 1$ and dark red color for $\big<\ell_{z}\big>\approx -1$). Following the semi-classical argument of Ref. \cite{Sinova}, it is possible to show that $\sigma_{OH}^z$ is a consequence of the existence of non-trivial in-plane orbital texture ($\big<\ell_{x,y}\big>_{n,\vec{k}}^s$). Some features of the function $\sigma_{OH}^z(E_f)$ can be understood from these textures, as we briefly mentioned in the main text, and now we detailed here.

\begin{figure}[H]
	\centering
	\includegraphics[width=1.\columnwidth]{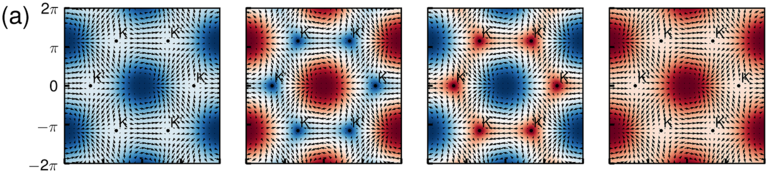}
	\includegraphics[width=1.\columnwidth]{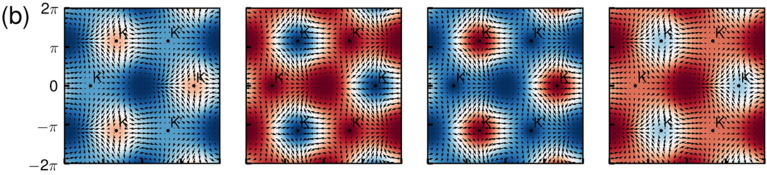}
	\includegraphics[width=1.01\columnwidth]{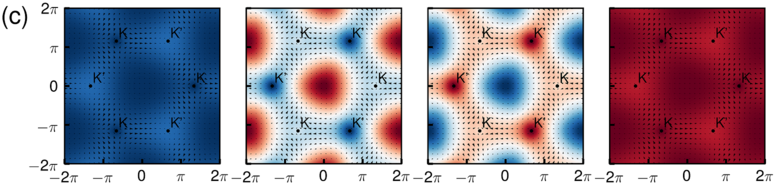}
	\caption{ Orbital character of the $\uparrow$-spin eigenstates of $H$ [Eq. \ref{UHhopping} or Eq. (1) of the main text] calculated for: (a) $\lambda_{I}=0.2V_{pp\sigma}$, and $V_{AB}=0$; (b) $\lambda_{I}=0.2V_{pp\sigma}$, and $V_{AB}=0.8V_{pp\sigma}$; (c) $\lambda_{I}=1.1V_{pp\sigma}$, and $V_{AB}=0.8V_{pp\sigma}$.}
	\label{fig:SMtextures}
\end{figure}

Figure 2 of the main text displays both the in-plane and the out-of-plane orbital polarisations of the lowest $\uparrow$-spin energy band for the B1, A1 and B2 phases. Results for the $\downarrow$-spin bands can be easily obtained by time-reversal symmetry operation. In Figure \ref{fig:SMtextures} we complement our analysis by showing the orbital textures of the four $\uparrow$-spin energy bands for each one of the three phases. The orbital projections depicted in panel (a) were calculated for $\lambda_{I}=0.2V_{pp\sigma}$ and $V_{AB}=0$, and correspond to the case in which the system assumes the B1 phase.
Clearly, the in-plane orbital textures of the first and second energy bands are opposite to each other, and the same happens to the third and fourth bands, which leads $\sigma_{OH}^z(E_f)$ to be an odd function of Fermi energy, and consequently, the absence of OHE in the central gap. As it was shown in Fig. 3 of the main text, if we include second neighbors hopping in the tight-binding Hamiltonian, the particle-hole symmetry around the central gap is broken, and the cancelation of in-plane orbital texture is lost, leading to the appearance of a central plateau in the orbital Hall conductivity. It is also noteworthy that $\left<\ell^z\right>^{\uparrow}_{n,\vec{k}}$ for the second and third bands are opposite, as well as around the $K$($K'$) and $\Gamma$ symmetry points. Conversely, the first and fourth bands respectively exhibit $\left<\ell^z\right>^{\uparrow}_{n,\vec{k}}\approx \mp1$ in the vicinities of the $\Gamma$ point, but virtually vanishing values around $K$ and $K'$. Panel (b) displays the orbital projections of the eigenstates corresponding to the A1 phase, calculated for $\lambda_{I}=0.2V_{pp\sigma}$ and $V_{AB}=0.8V_{pp\sigma}$. One of the main eye-catching characteristics of this phase is the opposed out-of-plane orbital polarisations around the $K'$ and $K$ points, which is a manifestation of the orbital-valley locking produced by $V_{AB}$. Similarly to phase B1, the out-of-plane polarisations of the first and second $\uparrow$-spin energy bands are opposed to the fourth and third ones, respectively. In addition, the in-plane orbital angular momentum polarisations for this phase exhibit the same configuration as those obtained for the B1 phase,which means that, also in this phase, sigma is an odd function of Fermi energy, with no central plateau. However, due to the orbital-valley locking, the corresponding absolute values are smaller, which explains the different curve derivative of the OHE in the phase A1 when compared with the OHE of the phase B1. Finally, panel (c) shows the orbital character of the system, calculated for $\lambda_{I}=1.1V_{pp\sigma}$ and $V_{AB}=0.8V_{pp\sigma}$, when it is in the B2 phase. In this case we find that $\left<\ell^z\right>^{\uparrow}_{n,\vec{k}}\approx -1$ for the lowest energy band, which goes along with a substantial reduction of the in-plane texture. Similarly to the previous cases, $\left<\ell^z\right>^{\uparrow}_{n,\vec{k}}$ for the lowest and highest energy bands are inverted. However, there a noticeable change in $\left<\ell^z\right>^{\uparrow}_{n,\vec{k}}$ in comparison with the results obtained for the A1 phase, which is accompanied by a relatively strong orbital-valley locking produced by the combined action of the large values of $\lambda_{I}$ and $V_{AB}$.

\section{Low-energy approximation}

\begin{figure}[h]
	\centering
	\includegraphics[width=0.65\columnwidth]{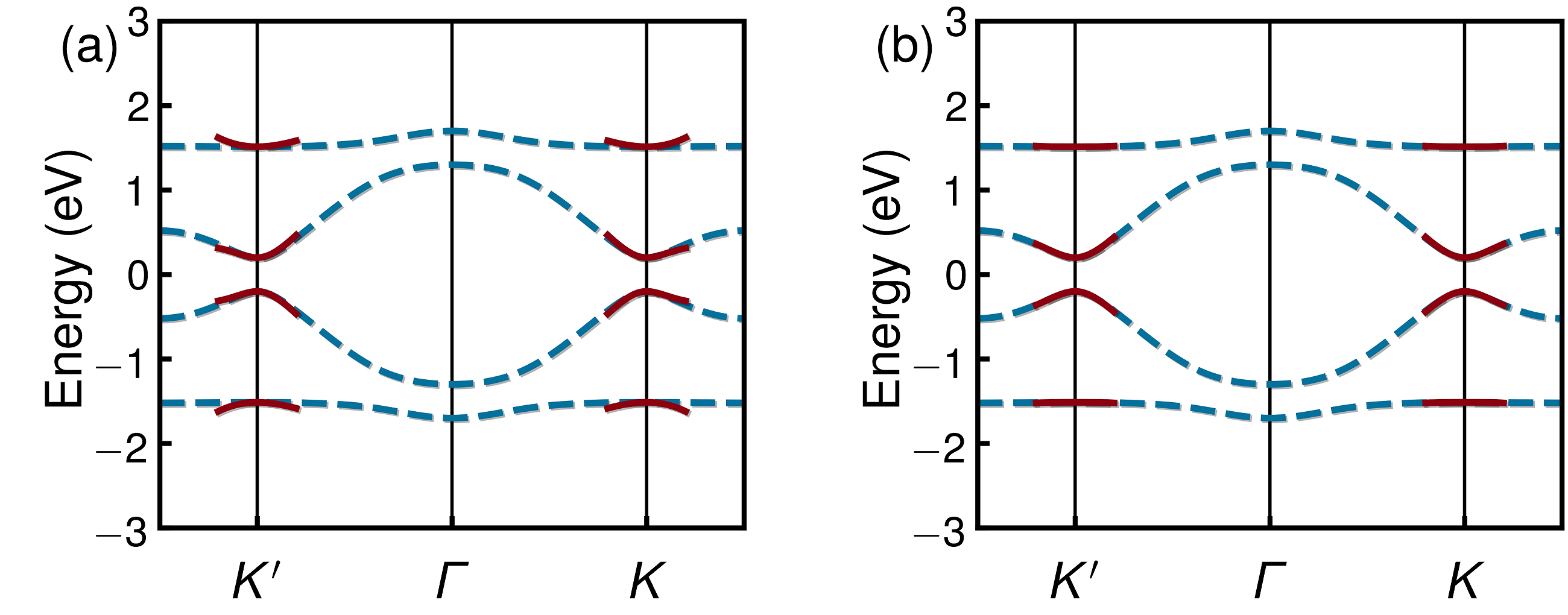}
	\includegraphics[width=0.65\columnwidth]{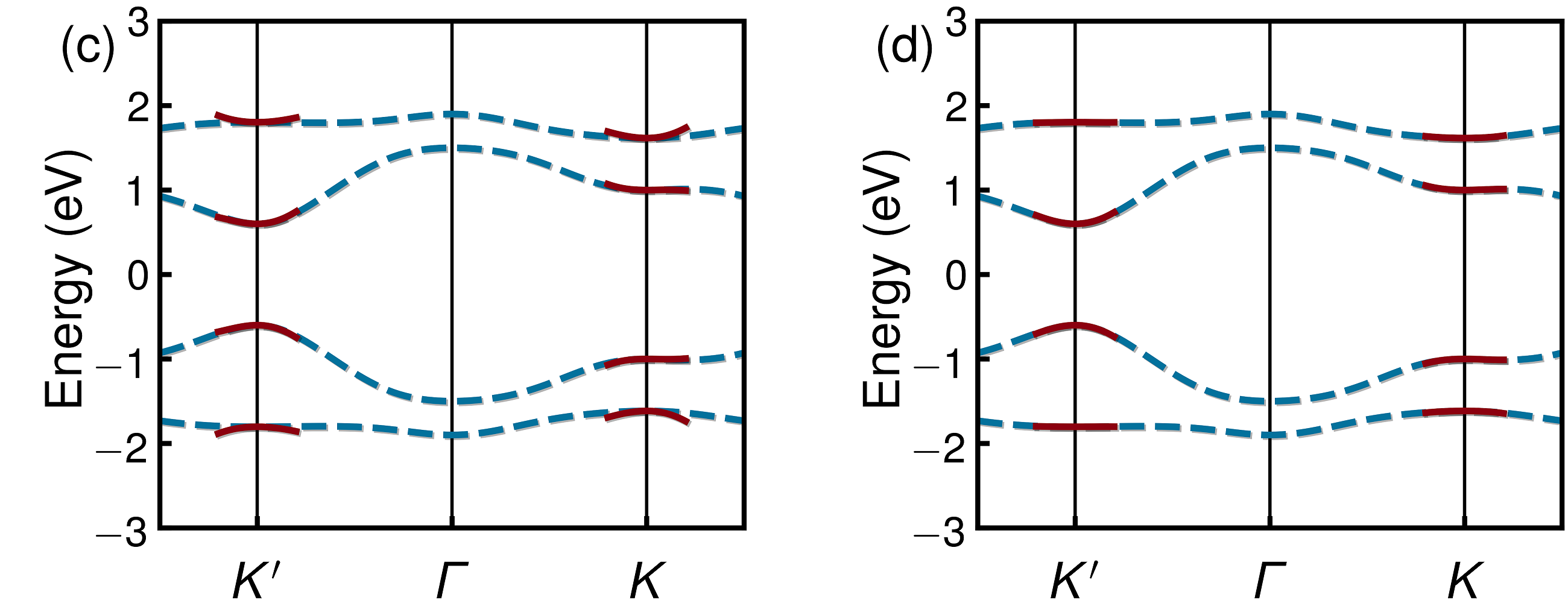}
	\includegraphics[width=0.65\columnwidth]{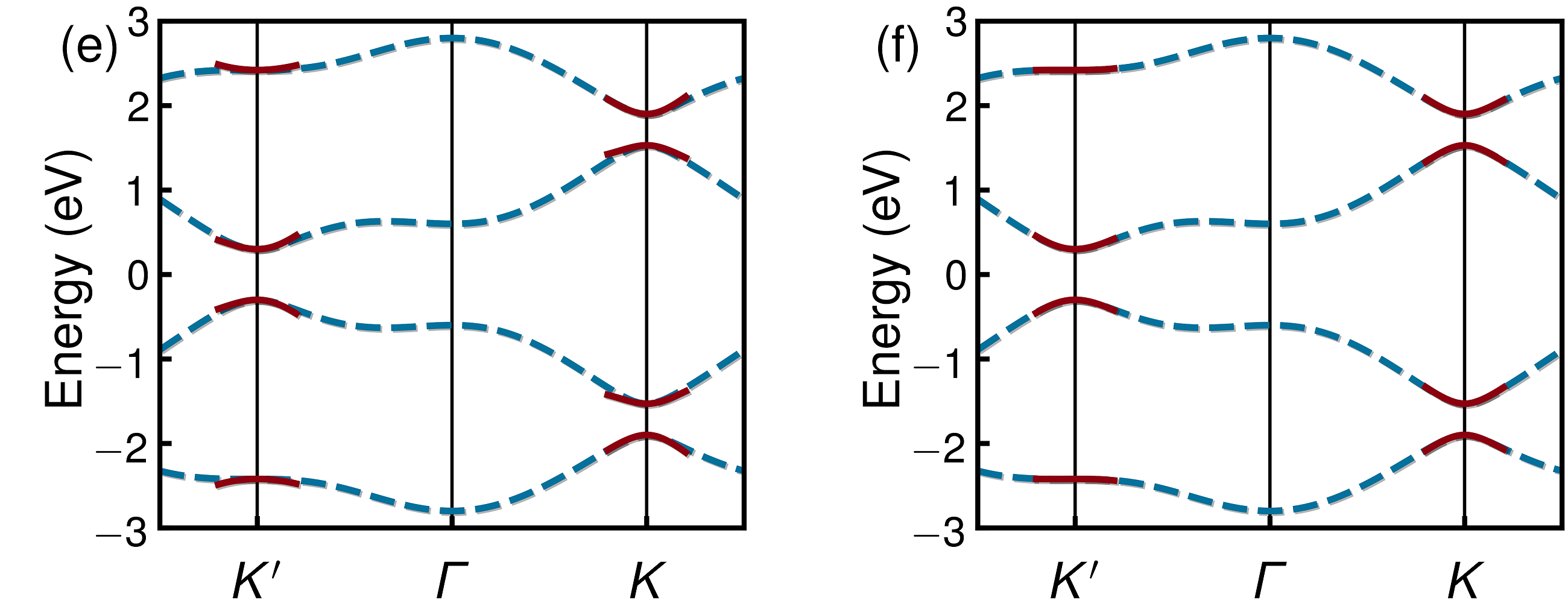}
	\caption{\label{fig:compare}
		Comparison between the tight-binding energy band calculations (blue dashed lines) with the eigenvalues of our effective Hamiltonians in the vicinities of $K$ and $K' $ (red solid lines). The eigenvalues obtained with the linear and quadratic order expansions are depicted in the left and right panels, respectively. The results are for: (a) $\lambda_{I}=0.2V_{pp\sigma}$ $V_{AB}=0$, (b) $\lambda_{I}=0.2V_{pp\sigma}$ $V_{AB}=0.8V_{pp\sigma}$, and (c) $\lambda_{I}=1.1V_{pp\sigma}$ $V_{AB}=0.8V_{pp\sigma}$. }
\end{figure}

As discussed in the main text, our effective Hamiltonian in the vicinity of the $K/K'$ point can be expressed in terms of $SU(2) \otimes SU(2)$ orbital and sub-lattice algebras. Expanding the matrix of Eq. \ref{UHhopping} near valleys $K=4\pi/3a$ and $K'=-4\pi/3a$, we obtain, up to first order in electronic momentum, the effective theory 

\begin{align}
\mathcal{H}_{\text{eff}} = -\hbar v_F \left(k_x\sigma_x+\tau k_y\sigma_y\right)+s\lambda_{I}\ell^z +V_{AB}\sigma^{z}+H_{\ell}.
\label{eqn:Heff}
\end{align}
Here, $v_{F} = \frac{a\sqrt{3}}{2\hbar}V_{pp\sigma}$ represents the Fermi velocity, and $a$ is the lattice constant; $\tau = \pm1$ for the $K$ and $K'$ valleys, respectively, and $s=\pm1$ for $\uparrow$ and $\downarrow$ spin electrons, respectively. The last term $H_{{\ell}}$ breaks the degeneracy between $\ell^z$ eigenstates and can be separated in two contributions:

\begin{align}
H_{\ell} = H_{\ell k}+H_{D}, ~\mbox{where}~~  H_{\ell k}=-\frac{\hbar v_{F}}{4}\tau\left( k_{+} \ell_{+}\sigma_{\tau} + k_{-}\ell_{-}\sigma_{\bar{\tau}} \right) ~~\mbox{and}~~ H_D=-\frac{\sqrt{3}\hbar v_{F}}{2a}\left ( \ell_{x}\sigma_{x}+\tau \ell_{y}\sigma_{y}\right).
\label{eqn:SMOrbitalTerm}
\end{align} 
\noindent
$\sigma_\tau=\sigma_{x}+ i\tau\sigma_{y}$, $\bar{\tau}=-\tau$, $\ell_{\alpha}$ ($\alpha = x,y$) represent the orbital angular momentum matrices in the corresponding Hilbert space, $k_{\pm}=k_x\pm i k_y$, and $\ell_{\pm}=\ell_x\pm i \ell_y$.

\begin{figure}[h]
	\centering
	\includegraphics[width=0.3\columnwidth]{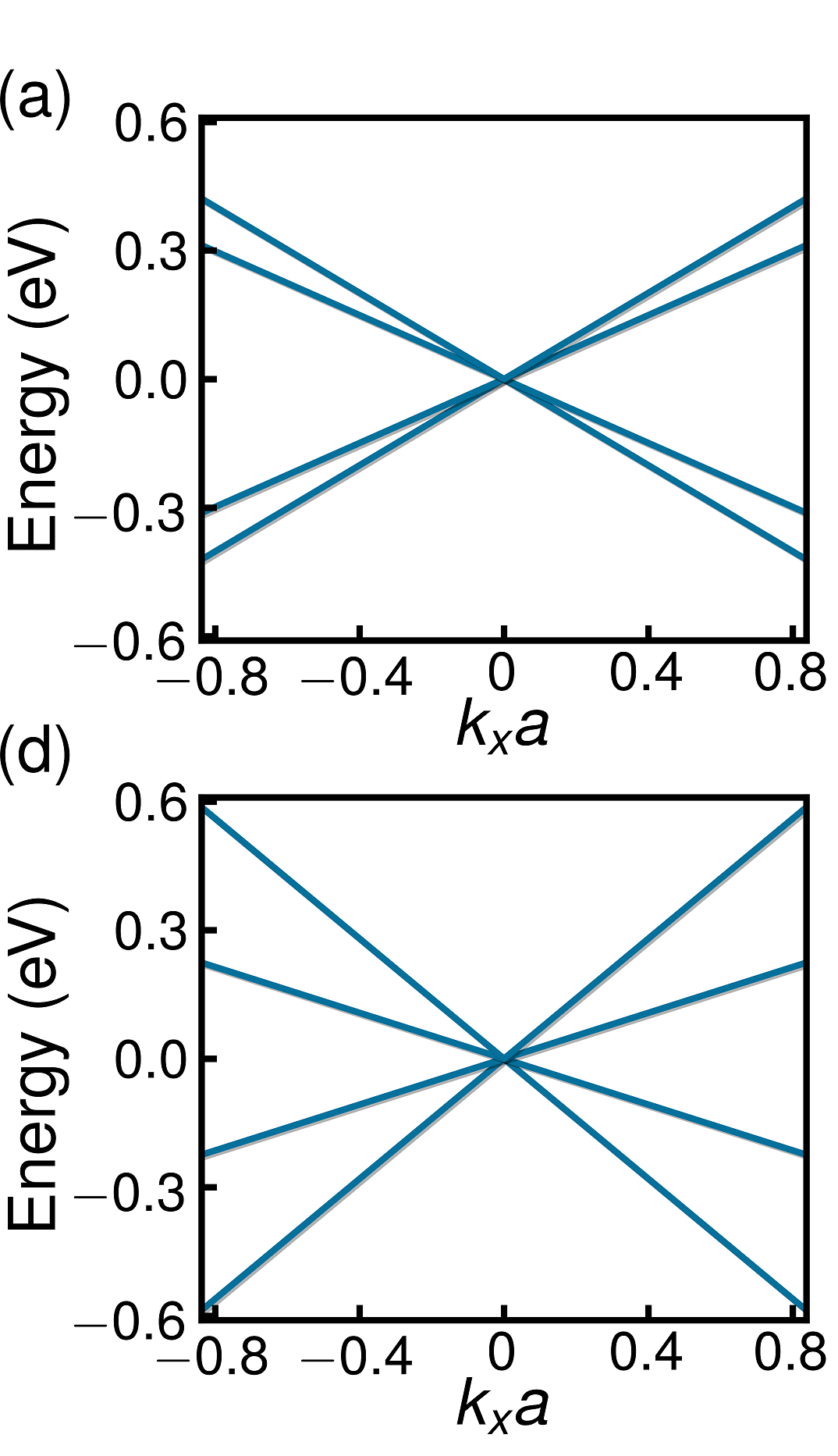}
	\includegraphics[width=0.3\columnwidth]{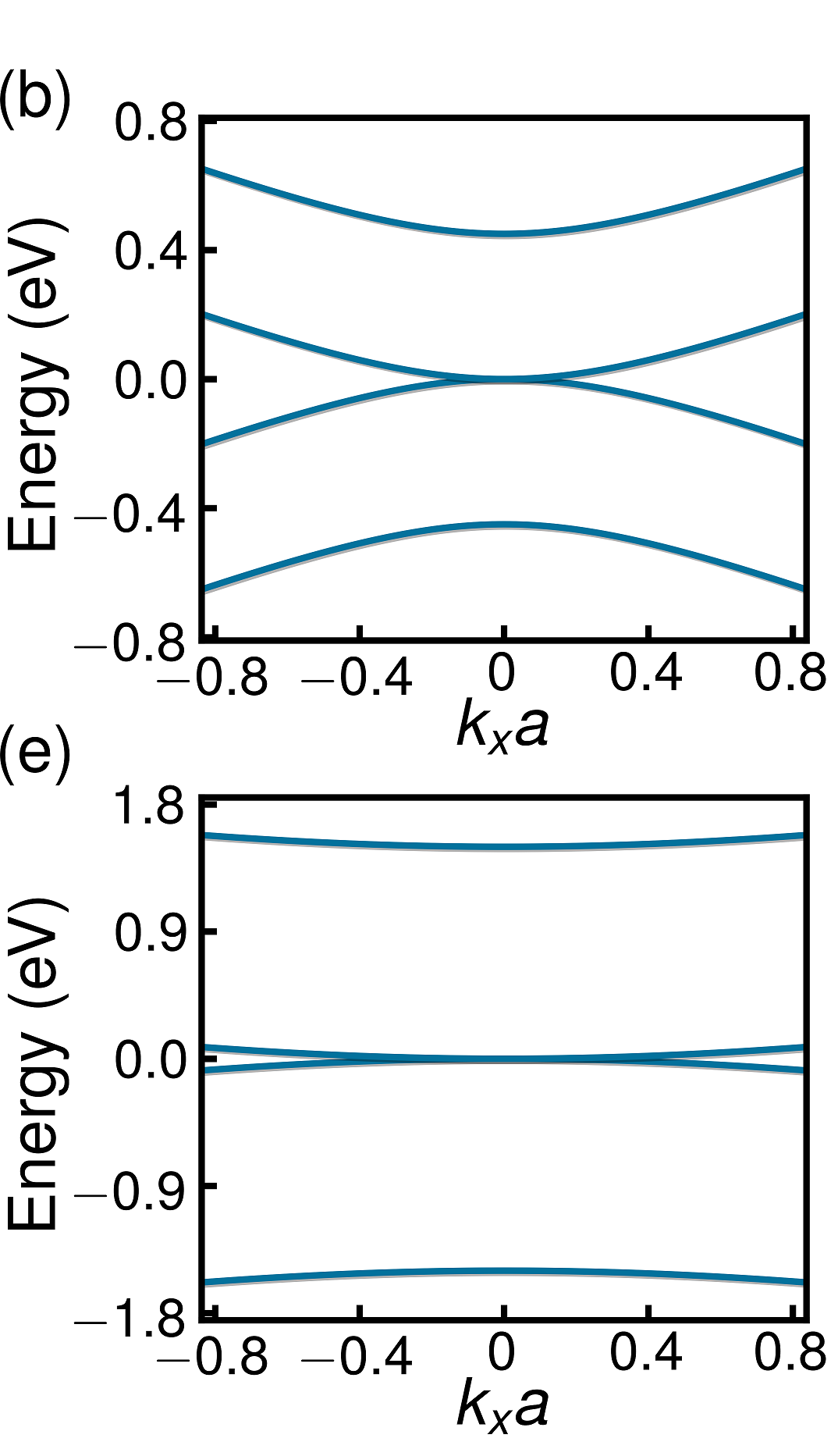}
	\includegraphics[width=0.3\columnwidth]{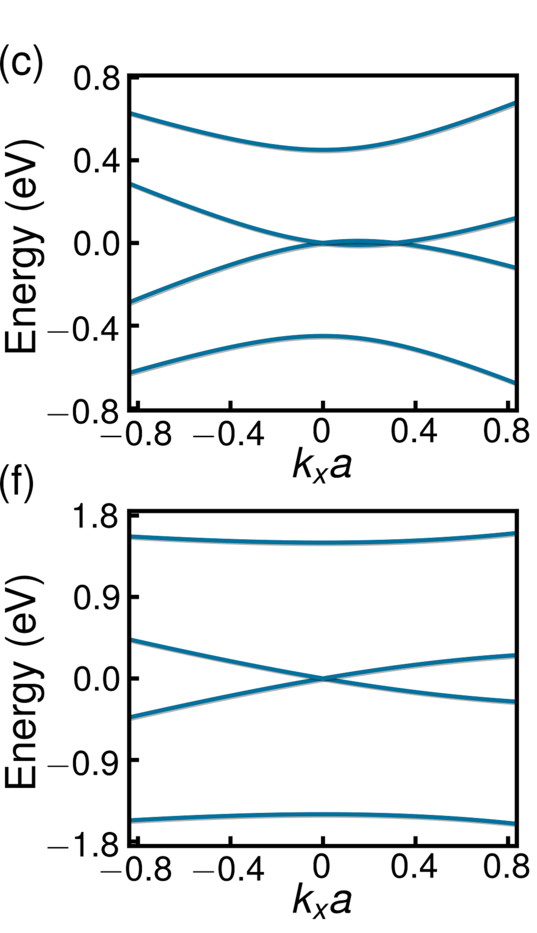}
	\caption{\label{fig:Hell}Energy bands calculated as functions of $k_x$ (for $k_y=0$) by means of our effective theory around the $K$ symmetry point. Panels (a), (b) and (c) depict the results obtained for $\eta=0.3$ in the cases: $H_{\ell K}\neq 0, H_D=0$; $H_{\ell K}=0, H_D\neq 0$ and $H_{\ell K}\neq 0, H_D\neq 0$, respectively. Panels (d), (e), and (f) show the results calculated for the same cases, but with $\eta=1.0$.}
\end{figure}

Figure \ref{fig:compare} shows a comparison between the energy band spectra obtained by our tight-binding (blue dashed lines) and effective models (red solid lines) calculations in the vicinities of $K$ and $K'$. In the left column we notice for the three phases that our effective linear model describes rather well the two inner energy bands, but fails to properly do so for the two outer ones. This can be corrected with the inclusion of quadratic terms in our approximation, as illustrated in the right column of Figure \ref{fig:compare}. It is noteworthy that the orbital texture near $K$ and $K'$ are very well described by our effective model. Nevertheless, to reproduce the orbital texture in the vicinity of $\Gamma$, it is necessary to perform an even higher-order expansion up to 4th order.\\

To provide insight on how $H_\ell$ affects the energy spectrum and orbital textures of this model, we examine the corresponding contributions of each term in Eq. \ref{eqn:SMOrbitalTerm}. For simplicity, we consider only one spin sector. In this case, the energy spectrum of $H_0 = -\hbar v_F \left(k_x\sigma_x+\tau k_y\sigma_y\right)$ consists of two degenerate Dirac cones that are associated with the two eigenstates of the angular momentum pseudo-spinor. Similarly to what occurs in graphene, the inclusion of $H_{AB} = V_{AB}\sigma^{z}$ opens an energy gap in the spectrum, while $H_{\text{SOC}} = s\lambda_{I}\ell^z $ acts as an orbital exchange interaction, shifting upwards (downwards) the Dirac cone associated with the $\ell_z$ eigenvalue +1(-1). To understand how $H_\ell$ modifies the energy spectrum, we introduce a multiplicative factor that regulates its overall intensity and inspect the energy band structure of $H_0+\eta H_\ell$ for two different values of $\eta$ in the following situations: (i) $H_{\ell K}\neq 0, H_D=0$, (ii) $H_{\ell K}=0, H_D\neq 0$, and (iii) $H_{\ell K}\neq 0, H_D\neq 0$. The results for the energy bands calculated as functions of $k_x$ for $k_y=0$ are exhibited in Figure \ref{fig:Hell}. In panels (a) and (d) we note that $H_{\ell K}$ lifts the orbital degeneracy of the two Dirac cones for $k_x \ne 0$, by differently renormalising their corresponding Fermi velocities. Panels (b) and (d) show how $H_D$ affects the energy bands. $H_D$ does not depend upon the wave vector $\vec{k}$, and has the same functional form of a Dresselhaus SOC for Dirac Fermions. It may be regarded as equivalent to a Dresselhaus SOC for orbital states. As expected, $H_D$ leads to a Dresselhaus-like band splitting, without opening a gap at $E=0$.  In panels (c) and (f) of Figure \ref{fig:Hell}, we clearly see the formation of a single Dirac cone and the two outer bands when both $H_{\ell K}$ and $H_D$ are present. It is worth recalling that to reproduce the flat-bands, it is necessary to consider high-order terms in $k$. Similarly to what is observed in quantum anomalous Hall insulators, the gap opening at $E=0$ is a consequence of the interplay between the orbital equivalent of a SOC and an exchange interaction. There is, however, a rich phenomenology involving the contributions of the distinct terms in Eq. \ref{eqn:Heff} that arises when $\eta$ is varied, but this goes beyond the scope of the present discussion. 

\begin{figure}[h]
	\centering
	\includegraphics[width=1.0\columnwidth]{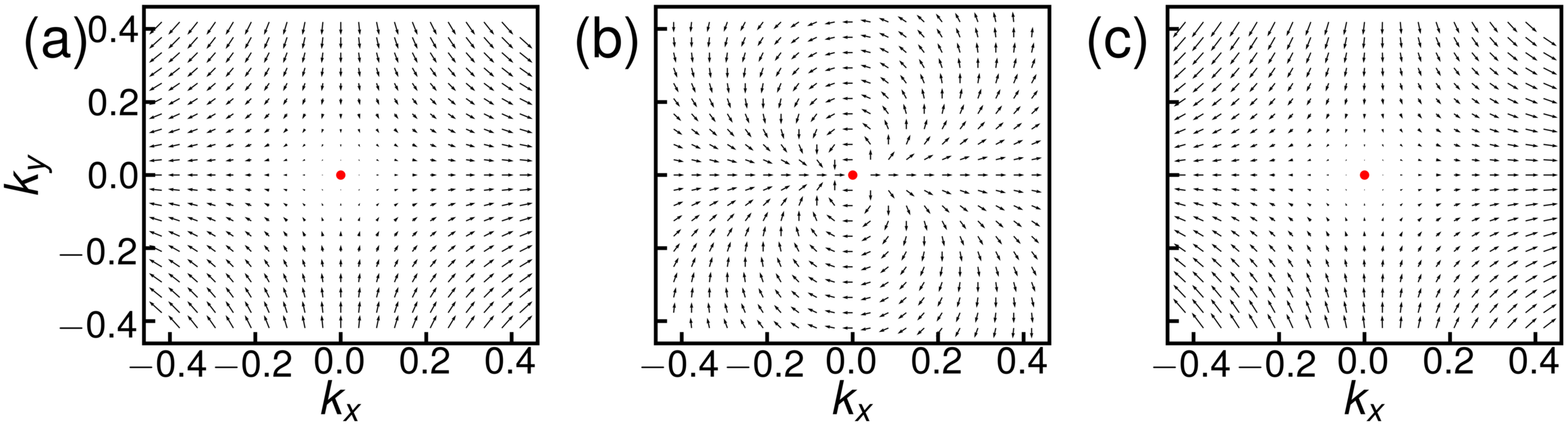}
	\caption{Comparison between of the in-plane texture profile for: (a) $H_D\neq 0$ and $H_{\ell K}= 0$; (b) $H_D= 0$ and $H_{\ell K}\neq0$, and (c) $H_D\neq 0$ and $H_{\ell K}\neq0$. \label{fig:comparatexture}}
\end{figure}
Finally, we examine the role of $H_\ell$ and $H_D$ in the orbital texture of this model system.  Figure \ref{fig:comparatexture} shows the orbital textures calculated for: (a) $H_D\neq 0$ and $H_{\ell K}= 0$;  (b) for $H_D= 0$ and $H_{\ell K}\neq$, and (c) for the effective complete Hamiltonian without SOC and $V_{AB}$. By comparing the three panels, it is clear that the orbital texture of our effective model is basically governed by the Dresselhaus-like coupling associated with the orbital angular momentum spinor, which reproduces rather well the in-plane texture of our tight-binding calculations near $K$.

\section{Second nearest neighbours and orbital texture analysis}

As mentioned in the main text, the absence of OHE plateau in the central electronic spectrum gap of px-py-model of Eq. (1) is a consequence of the combination of the particle-hole and parity symmetries of spectrum which translates in cancellation of in-plane orbital texture at half-filling. To understand better the consequences of the breaking of these symmetries, we introduce a toy-model of the $p_x$-$p_y$ Hamiltonian, where we have included second nearest-neighbours hopping. This model is described by,

\begin{equation}
{\cal H}=\sum_{\langle i j\rangle} \sum_{\mu \nu s} t_{i j}^{\mu \nu}{p^\dagger_{i \mu s}}p_{j \nu s}+\sum_{\langle\langle i j\rangle \rangle} \sum_{\mu \nu s} t_{i j}^{\mu \nu}{p^\dagger_{i \mu s}}p_{j \nu s}+ \sum_{i \mu s}\epsilon_{i} p^\dagger_{i \mu s} p_{i \mu s}+   \sum_{i \mu s} \mathbf{h}^z_{\mu s}  p^\dagger_{i \mu s} p_{i \mu s}, 
\label{eqn:H2NN}
\end{equation} 

\noindent here as before, $i$ and $j$ represents the honeycomb lattice sites whose position is given by $\vec{R}_i$ and $\vec{R}_j$, respectively. The symbols $\langle i j \rangle$ and $\langle \langle i j \rangle \rangle$ indicates that the summations are restricted to the nearest and second nearest neighbour sites respectively. The operator $p^{\dagger}_{i \mu s}$ creates an electron of spin $s$ in the atomic orbitals $p_\mu=p_{\pm}=\frac{1}{\sqrt{2}}(p_x\pm ip_y)$ centred at $\vec{R}_i$. Here, $s=\,\uparrow,\downarrow$ labels the two electronic spin states,  and, now, $\epsilon_i$ is the atomic energy at site $i$, which encodes the effect of the combination of a sublattice potential $V_{AB}$, and the on-site energy of $p$ orbitals  $\varepsilon_p$. This terms take values $\epsilon_i = \varepsilon_p\pm V_{AB}$, when site i belongs to the A and B sub-lattices of the honeycomb arrangement, respectively. \\ 

\begin{figure}[h]
	\centering
	\includegraphics[width=0.98\columnwidth]{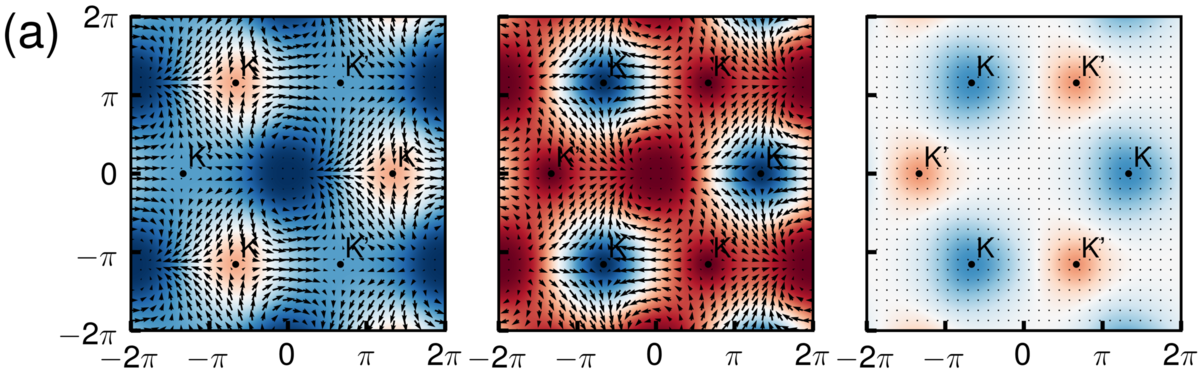}
	\includegraphics[width=0.98\columnwidth]{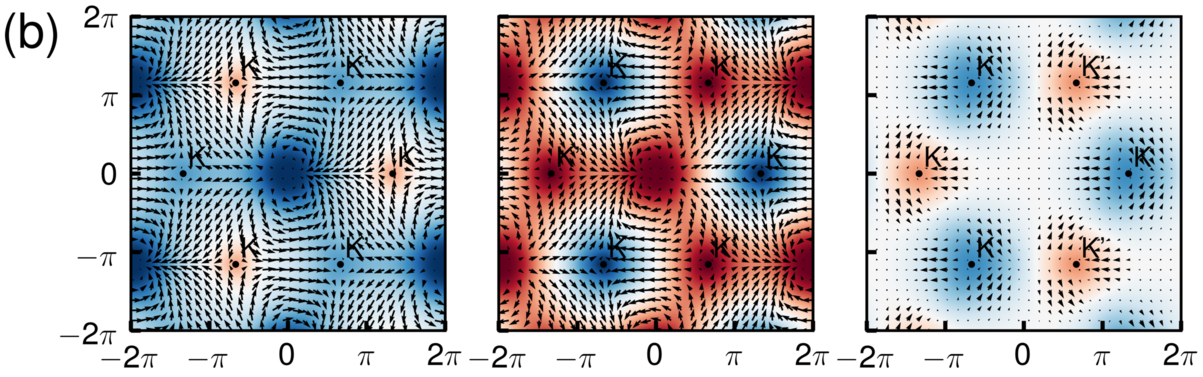}
	\caption{Comparison between the orbital (spin-$\uparrow$) texture profiles of the $p_x$-$p_y$ with only nearest neighbours (a), and the orbital texture of the same model when second nearest neighbours are considered (b). Left: Orbital Texture profile for the deepest energy band. Center: Orbital Texture profile for the second lowest energy band. Right: addition of the orbital textures in Left and Right panels with the in-plane component scaled by a factor $5$. }
	\label{fig:OrbitalTextures}
\end{figure}

As it was shown in figure 3 of the main text, the principal effect of the particle-hole and parity symmetries breaking, a consequence of the inclusion of the second nearest neighbours, is the appearance of an orbital Hall conductivity plateau in the central gap. In order to uncover the connection between the appearance of this plateau and the orbital textures, we analyse the texture profiles of the two deepest energy bands for two different cases of this model. In panel (a) of the figure \ref{fig:OrbitalTextures} are shown the orbital textures of the two deepest energy bands(left and central panels) of the simple model that does not include second nearest neighbours [Eq. (1) in the main text] and their summation (right panel) in which the in-plane components of the texture are in a larger scale to make easier the analysis of their details. The in-plane component of orbital textures in left and central panels present the dipole configuration around the $\Gamma$ point and the anti-vortices in the $K$ and $K'$ points, and the out-of-plane component appears due to the inversion symmetry breaking produced by the inter-lattice potential. At the right panel of the figure (a), we show that the addition of the orbital textures of the left and central panels results in a zero net in-plane orbital texture. Once that orbital Hall conductivity ($\sigma_{OH}^z$) appears as a consequence of dynamics of in-plane orbital texture, in presence of an external electric field, this explains the absence of OHE in the central gap of the simplified $p_x$-$p_y$ model of Eq. (1) in the main text. Now, in panel (b) of the figure \ref{fig:OrbitalTextures}, we consider the orbital texture of the Hamiltonian with the inclusion of second nearest-neighbours hopping (see. Eq. \ref{eqn:H2NN}). Again, the left and the central panels of the figure show the orbital textures of two deepest bands and the right panel shows the sum of these two textures, with the in-plane component multiplied by a scaling factor to facilitate its visualization. To maintain the resemblance between the aforementioned case and this new case, we set the same Slater-Koster parameters that we used for the phase $A1$ of the simplified model with the addition of $\varepsilon_p=-0.3$ and $V_{pp\sigma2}=-0.2$. With these parameters, as it was shown in figure 3 (a) of the main text the energy bands of this modified model are not particle-hole symmetric, an effect caused exclusively by the inclusion of second nearest-neighbours. We note in Fig. \ref{fig:OrbitalTextures}  (b) that the overall features of in-plane orbital-texture of two deepest bands (left and central panels) are not qualitatively modified, i.e., still present a dipole-like texture near $\Gamma$-point and anti-vortices textures at valleys. However, as can be seen in Fig. \ref{fig:OrbitalTextures} (b), right panel, the exact cancellation of the in-plane texture of two deepest bands is lost, causing the existence of a net in-plane orbital texture which produces an OHE in the central gap of the spectrum shown in figure 3 (b) of the main text.  \\

Once shown by means of the simply $p_x-p_y$ model that the orbital Hall effect is present in systems where the particle-hole and parity symmetries are absent,  we now focus on a real material. For this purpose, we have chosen the flat bismuthene grown on SiC as a candidate for the observation of OHE in the central plateau. The observation of orbital-insulator phase in the central gap of bismuthene should be easier in the experimental point of view, once it corresponds to neutrality situation. In the past, this system has been studied by means of the aforementioned minimal $p_x$-$p_y$ tight-binding Hamiltonian \cite{ReisBismutheneExperimental,FDominguez-TestingTOpologicalprotection,PRLPxPy-Nosotros}. However, we have noticed that by including second nearest-neighbours in the tight-binding Hamiltonian used in Ref. \onlinecite{PRLPxPy-Nosotros} the electronic structure is better reproduced. In the bismuthene/SiC heterostructure, the break of inversion symmetry induces a small Rashba SOC,

\begin{equation}
H_R  = 2i\lambda_{R}\sum_{\langle i,j \rangle}\sum_{\mu \nu s}
p^\dagger_{i \mu \bar{s}} \left[\hat{z}\cdot\left(\vec{\sigma}\times\hat{e}_{ij}\right)\right]_{\bar{s} s}p_{j \nu s} + H.c.
\label{eqn:rashba}
\end{equation}

\noindent where $\vec{\sigma}$ symbolises the Pauli vector, $\hat{e}_{ij}$ denotes the unit vector along the n.n. inter-site direction of $\vec{R}_j - \vec{R}_i$,  $\lambda_{R}$ is the Rashba SOC constant, and $\bar{s}$ designates the opposite spin direction specified by $s$. We will consider this term only in the fitting of tight-binding Hamiltonian to DFT spectrum and, we neglect it in transport calculations presented here. The reasons is that the non-conserving spin character of this coupling complicates the analysis of orbital texture and its typical small value does not alter the main conclusions of our discussion, as it was checked by us. In figure \ref{fig:bismuthenetb} is shown a direct comparison of the DFT energy band structure obtained in Ref. ~\onlinecite{ReisBismutheneExperimental},  with and without the inclusion of Rashba spin-orbit coupling. From this figure, it is rapidly noticeable the agreement between the DFT energy bands and the tight-binding model in describing the top of the valence band and the bottom of the conduction bands and the indirect gap in the $\Gamma$ point. In the table \ref{tab:parametrosSKbismuthene2NN} are shown the two-centre integral parameters used in the description of this model. Once shown the agreement of energy band of the complete model, we are going to restrict ourselves to the situation in which the Rashba SOC is neglected and the system is subject to a staggered potential $V_{AB}=0.87$. The first of these constraints is to avoid complications in the analysis due to possible contributions to the orbital texture by the Rashba SOC, which does not conserve spin as a good quantum number, and the second one is to leave the system in a topological phase similar to the phase $A1$ of the $p_x-p_y$ model with only nearest neighbours. This will allow us to focus on the analysis of the Orbital texture and its connection with the orbital Hall conductivity.

\begin{figure}[h]
	\centering
	\includegraphics[width=0.98\linewidth]{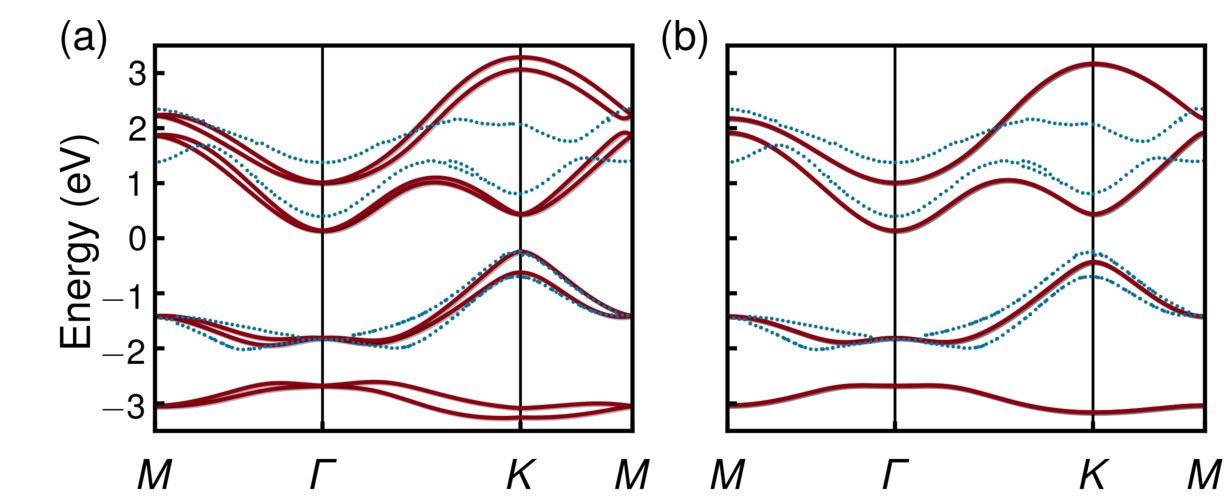}
	\caption{Comparison between the DFT energy bands (blue doted line) and the tight-binding model with second nearest neighbours bands (red solid line) for: (a) $\lambda_{R}=0.032$ eV  and (b) $\lambda_{R}= 0$.}\label{fig:bismuthenetb}
\end{figure}

\begin{table}[h] 
	\centering
	
	\caption{Second Nearest-neighbour two-centre energy integrals, and spin orbit coupling constants (all in eV) for the Bi/SiC.}
	\label{tab:parametrosSKbismuthene2NN}
	\begin{tabular}{l c c c}
		\hline	
		Two-centre integrals  & Intrinsic SOC & Rashba SOC & On-site energy \\
		\hline
		$V_{pp\sigma}=+1.51522$ & $\lambda_{I} = -0.435$ & $\lambda_{R}=0.032$ &$\epsilon_{p} = -0.279865$ \\
		$V_{pp\pi} = -0.575788$ &{}& {} & {}\\
		$V_{pp\sigma2}=-0.18$ & {}&{} &{}\\
		$V_{pp\pi2} = -0.00658$ & {} & {} & {}\\

	\end{tabular}
\end{table}

In figure \ref{fig:bismutheneOrbtex} is displayed the orbital textures of the two deepest energy bands of bismuthene grown over SiC in the phase $A1$ and without Rashba SOC. The principal difference that is noticeable when one looks at this figure is the change of the out-of-plane orbital texture of the spin-$\uparrow$ sector, with respect to textures of $A_1$-phase in previous cases, produced by the change of sign of the spin-orbit coupling. The in-plane components of orbital texture for the model with parameters of bismuthene shown in Fig. \ref{fig:bismutheneOrbtex} left and centre panels, do not present noticeable differences from those of Fig. \ref{fig:OrbitalTextures} (b) of the model with second nearest neighbours. Again, as can be seen from the right panel of Fig. \ref{fig:bismutheneOrbtex}, there is a non-zero total in-plane orbital texture when we add up the textures of two deepest bands of left and central panels what again explain the existence of OHE in the central gap, as shown in Figure 4 of the main text. This suggests the recent synthesized flat bismuthene as a realistic platform to observe the orbital Hall insulator phase. The central plateau will persist by the inclusion of the Rashba term [Eq. (\ref{eqn:rashba})] on Hamiltonian, once the spectrum keeps the particle-hole asymmetry. When the Rashba term is included, we cannot separate the textures by spin sectors because it breaks the $s_z$-symmetry. So the previous analysis of sum of orbital texture must be done summing the four lowest energy bands to obtain total texture related to the central plateau. But the main conclusions are the same and we do not present this analysis here.

\begin{figure}[h]
	\centering
	\includegraphics[width=0.98\columnwidth]{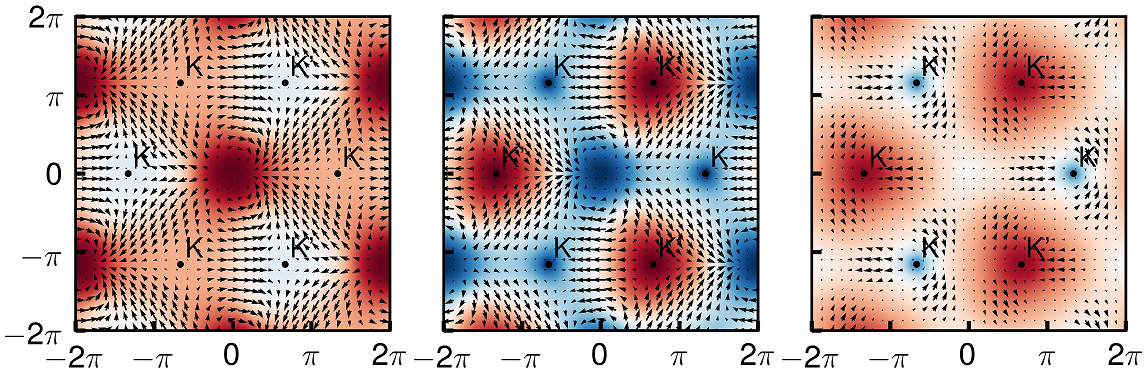}
	\caption{Orbital Textures profile of spin-$\uparrow$ sector for  the two lowest energy bands of bismuthene over SiC in which the Rashba SOC is neglected and $V_{AB}=0.87$ eV. Left: Orbital texture profile of the lowest energy band. Center: Orbital texture profile of the second lowest energy band. Right: Addition of the later texture profiles. The resultant in-plane orbital texture are scaled by a factor $5$ to facilitate the visualization}
	\label{fig:bismutheneOrbtex}
\end{figure}
\end{widetext}

\end{document}